\documentclass[twocolumn,10pt]{IEEEtran}

\usepackage{amssymb}
\usepackage{amsmath}
\usepackage{amsthm}
\usepackage{graphicx}
\usepackage{setspace}
\usepackage[linesnumbered,ruled]{algorithm2e}
\usepackage{comment}
\usepackage{xcolor}
\usepackage{lscape}
\usepackage{arydshln}
\usepackage{booktabs}
\usepackage{cancel}
\usepackage{subfigure}

\newtheorem{proposition}{Proposition}

\usepackage[acronym]{glossaries}
\newacronym{awgn}{AWGN}{additive white Gaussian noise}
\newacronym{siso}{SISO}{single-input single-output}
\newacronym{star-ris}{STAR-RIS}{simultaneously transmitting and reflecting RIS}
\newacronym{ios}{IOS}{intelligent omni-surface}
\newacronym{rsma}{RSMA}{rate splitting multiple access}
\newacronym{dfrc}{DFRC}{dual-function radar-communication}
\newacronym{csi}{CSI}{channel state information}
\newacronym{mmwave}{mmWave}{millimeter wave}
\newacronym{miso}{MISO}{multiple-input single-output}
\newacronym{mrt}{MRT}{maximum ratio transmission}
\newacronym{iid}{i.i.d.}{independent and identically distributed}

\newacronym{los}{LoS}{line-of-sight}
\newacronym{mimo}{MIMO}{multiple-input multiple-output}
\newacronym{nlos}{NLoS}{non-line-of-sight}
\newacronym{snr}{SNR}{signal-to-noise ratio}
\newacronym{upa}{UPA}{uniform planar array}
\newacronym{ula}{ULA}{uniform linear array}
\newacronym{cscg}{CSCG}{circularly symmetric complex Gaussian}
\newacronym{rf}{RF}{radio frequency}
\newacronym{sinr}{SINR}{signal-to-interference-plus-noise ratio}
\newacronym{ofdm}{OFDM}{orthogonal frequency division multiplexing}
\newacronym{noma}{NOMA}{non-orthogonal multiple access}
\newacronym{wpt}{WPT}{wireless power transfer}
\newacronym{swipt}{SWIPT}{simultaneous wireless information and power transfer}
\newacronym{bcd}{BCD}{block coordinate descent}

\newacronym{em}{EM}{electromagnetic}
\newacronym{sdr}{SDR}{semidefinite relaxation}
\newacronym{sdp}{SDP}{semidefinite programming}
\newacronym{bb}{B\&B}{branch-and-bound}
\newacronym{mm}{MM}{majorization-minimization}
\newacronym{admm}{ADMM}{alternating direction method of multipliers}
\newacronym{pin}{PIN}{positive-intrinsic negative}
\newacronym{vna}{VNA}{vector network analyzer}

\begin{document}

\title{A Universal Framework for Multiport Network Analysis of Reconfigurable Intelligent Surfaces}

\author{Matteo~Nerini,~\IEEEmembership{Graduate Student Member,~IEEE},
        Shanpu~Shen,~\IEEEmembership{Senior Member,~IEEE},\\
        Hongyu~Li,~\IEEEmembership{Graduate Student Member,~IEEE},
        Marco~Di~Renzo,~\IEEEmembership{Fellow,~IEEE},
        Bruno~Clerckx,~\IEEEmembership{Fellow,~IEEE}
        
\thanks{This work has been partially supported by UKRI grant EP/Y004086/1, EP/X040569/1, EP/Y037197/1, EP/X04047X/1, EP/Y037243/1.}
\thanks{The work of Marco Di Renzo was supported in part by the European Commission through the Horizon Europe project titled COVER under grant agreement number 101086228, the Horizon Europe project titled UNITE under grant agreement number 101129618, and the Horizon Europe project titled INSTINCT under grant agreement number 101139161, as well as by the Agence Nationale de la Recherche (ANR) through the France 2030 project titled ANR-PEPR Networks of the Future under grant agreement NF-YACARI 22-PEFT-0005, and by the CHIST-ERA project titled PASSIONATE under grant agreement CHIST-ERA-22-WAI-04 through ANR-23-CHR4-0003-01.}
\thanks{Matteo Nerini, Hongyu Li, and Bruno Clerckx are with the Department of Electrical and Electronic Engineering, Imperial College London, London SW7 2AZ, U.K. (e-mail: \{m.nerini20, c.li21, b.clerckx\}@imperial.ac.uk).}
\thanks{Shanpu Shen is with the Department of Electrical Engineering and Electronics, University of Liverpool, Liverpool L69 3GJ, U.K. (email: Shanpu.Shen@liverpool.ac.uk).}
\thanks{Marco Di Renzo is with the Universit\'e Paris-Saclay, CNRS, CentraleSup\'elec, Laboratoire des Signaux et Syst\`emes, 91192 Gif-sur-Yvette, France (e-mail: marco.di-renzo@universite-paris-saclay.fr).}}

\maketitle

\begin{abstract}
Reconfigurable intelligent surface (RIS) is an emerging paradigm able to control the propagation environment in wireless systems.
Most of the research on RIS has been dedicated to system optimization and, with the advent of beyond diagonal RIS (BD-RIS), to RIS architecture design.
However, developing general and unified \gls{em}-consistent models for RIS-aided systems remains an open problem.
In this study, we propose a universal framework for the multiport network analysis of RIS-aided systems.
With our framework, we model RIS-aided systems and RIS architectures through impedance, admittance, and scattering parameter analysis.
Based on these analyses, three equivalent models are derived accounting for the effects of impedance mismatching and mutual coupling.
The three models are then simplified by assuming large transmission distances, perfect matching, and no mutual coupling to understand the role of the RIS in the communication model.
The derived simplified models are consistent with the typical model used in related literature, although we show that an additional approximation is commonly considered in the literature.
We discuss the benefits of each analysis in characterizing and optimizing the RIS and how to select the most suitable parameters according to the needs.
Numerical results provide additional evidence of the equivalence of the three analyses.
\end{abstract}

\glsresetall


\section{Introduction}
\label{sec:intro}

Reconfigurable intelligent surface (RIS) is a promising technology expected to revolutionize future wireless systems \cite{wu21}.
An RIS is a surface made of multiple scattering elements, each of which can be reconfigured to impose an adaptive phase shift to the incident \gls{em} waves.
In this way, an RIS can dynamically manipulate the \gls{em} properties of the propagation environment, and enable the so-called concept of smart radio environment \cite{dir20}.
Since an RIS can manipulate the incident signal in a nearly passive manner, it is also a cost-effective solution characterized by low power consumption.

A substantial number of studies on RIS have been devoted to system optimization.
Specifically, RIS has been optimized to enhance the performance of wireless communication systems, including single-cell \cite{wu19b,guo20}, multi-cell \cite{pan20}, multi-RIS \cite{zhe21}, and wideband communications \cite{li21}.
Meanwhile, RIS has been integrated with multiple access schemes such as \gls{noma} \cite{xiu21} and \gls{rsma} \cite{ban21}, to ease the requirement of complex signal processing at the transmitter.
In addition to wireless communication systems, RIS has been also applied to \gls{wpt} systems \cite{fen22}, \gls{swipt} systems \cite{zha22a}, \gls{rf} sensing systems \cite{hu20}, and \gls{dfrc} systems \cite{liu22}.
While the aforementioned works \cite{wu19b}-\cite{liu22} apply an idealized RIS model, research has been conducted to optimize RIS in the presence of discretized reflection coefficients through low-complexity codebook designs \cite{di20,an22}, imperfect \gls{csi} \cite{che22}, and mutual coupling \cite{qia21,abr21,mur23,has23,akr23,abr23,pet23}.
Furthermore, RIS has been prototyped in \cite{dai20,zha21,rao22}, demonstrating its practicability.

The authors of \cite{wu19b}-\cite{rao22} have considered the optimization of conventional RIS architectures.
In a conventional RIS architecture, also known as single-connected RIS, the reflection coefficient of each RIS element is individually controlled through a tunable load, resulting in a diagonal scattering matrix \cite{she20}.
Recently, a novel advance in RIS, namely beyond diagonal RIS (BD-RIS), has been proposed \cite{li23-1}, which relies on novel RIS architectures yielding scattering matrices not limited to being diagonal.
The conventional RIS architecture has been first generalized by interconnecting groups of/all the RIS elements to each other through tunable impedance components, resulting in the group-/fully-connected architectures, respectively \cite{she20}.
Group- and fully-connected RIS have been optimized assuming continuous-valued impedances \cite{ner22,fan23}, considering discrete-valued impedances \cite{ner21}, and in the presence of mutual coupling \cite{li23-3}.
Forest-connected and tree-connected RIS have been proposed in \cite{ner23-1} to achieve the same performance enhancement as group- and fully-connected RIS, respectively, but at a reduced circuit complexity.
Furthermore, the RIS architectures achieving the best trade-off between performance and circuit complexity have been characterized in \cite{ner23-2}.
BD-RIS architectures have been investigated also with the objective of achieving full-space coverage, by reflecting and transmitting the incident signal \cite{li22-1}.
To improve the performance while preserving full-space coverage, multi-sector BD-RIS has been proposed \cite{li22-2,li23-2,wan23}.
Different from the fixed BD-RIS architectures studied in \cite{she20}-\cite{wan23}, dynamic BD-RIS architectures have been studied in \cite{li22,li22-3}, where higher flexibility is achieved by reconfiguring the interconnections among the RIS elements on a per channel realization basis.
An additional RIS architecture with a non-diagonal response is the so-called stacked intelligent metasurface (SIM), recently proposed to offer additional flexibility by staking multiple single-connected RISs \cite{an23}, whose communication model has been analyzed in \cite{ner24}.

While most of the research on conventional RIS and BD-RIS has primarily focused on RIS optimization \cite{wu19b}-\cite{rao22} and on RIS architecture development \cite{she20}-\cite{ner24}, only a few works have been devoted to model and analyze RIS-aided communication systems accounting for the \gls{em} properties of the RIS, such as impedance mismatching and mutual coupling \cite{dir22}.
%
With a focus on RIS modeled as an antenna array connected to a reconfigurable impedance network, three equivalent analyses are available to show the role of RIS in wireless networks from different perspectives.
First, in \cite{she20}, an RIS-aided communication model has been derived using multiport network analysis based on scattering parameters.
Second, in \cite{gra21}, another RIS-aided communication model has been developed by using the impedance parameters.
Interestingly, it has been shown in \cite{nos23} that the scattering parameters analysis of \cite{she20} and the impedance parameters analysis of \cite{gra21} lead to the same conclusion under the assumptions of perfect matching and no mutual coupling.
Third, another analysis based on admittance parameters has been proposed in \cite{ner23-1} to model BD-RIS with sparse interconnections among the RIS elements.
However, there is currently no universal framework for modeling RIS-aided systems according to impedance, admittance, and scattering parameters, as well as clarifying the meaning of the different components in each model, despite their equivalence being well-established in microwave theory.
Such a framework would improve our understanding of different \gls{em}-consistent models for RIS-aided systems, and would provide additional tools to perform RIS optimization.
Furthermore, it would enable a thorough analysis of the impact of each approximation and assumption necessary to simplify existing RIS-aided communication models based on the different parameters.

Motivated by the above considerations, in this paper, we move from existing works on \gls{em}-consistent RIS-aided communication models \cite{she20,gra21,nos23} and derive a universal framework to analyze RIS-aided communication systems.
This universal framework can be used to carry out a rigorous multiport network analysis of RIS-aided communication systems based on impedance, admittance, and scattering parameters, also denoted as $Z$-, $Y$-, and $S$-parameters, respectively.
As a result of these analyses, we can derive three RIS-aided communication models accounting for mutual coupling effects and impedance mismatching at the transmitter, RIS, and receiver.
Our universal framework shows the equivalence between channel models based on $Z$-, $Y$-, and $S$-parameters also under different assumptions and approximations.
This is achieved by providing the mappings between the terms of the models based on the three different parameters.
Moreover, we show that, under specific assumptions, the derived channel models boil down to models already used in previous literature, confirming the soundness and rigorousness of our analysis.
The contributions of this paper are as follows.

\textit{First}, we propose a universal framework to derive \gls{em}-consistent communication models for RIS-aided systems.
We exploit the proposed universal framework to analyze RIS-aided communication systems based on the $Z$-, $Y$-, and $S$-parameters.
Through these three independent analyses, we first derive three general channel models, accounting for the effects of impedance mismatching and mutual coupling at the transmitter, RIS, and receiver, which can significantly impact the performance of practical RIS designs.
Then, we show for the first time that the three derived models are equivalent when no approximations or assumptions are made.

\textit{Second}, since it is difficult to interpret the derived general channel models, we approximate them by considering the so-called unilateral approximation, applicable in the case of large transmission distances\footnote{The unilateral approximation consists in setting to zero the feedback channel between a transmitter and a receiver, i.e., the channel from the receiver to the transmitter, and holds when the electrical properties at the transmitter are independent of the electrical properties at the receiver \cite{ivr10}.}.
Under this approximation, we show that the three models are equivalent by deriving the mappings to convert the $Y$- and $S$-parameters into the $Z$-parameters.
Interestingly, the derived simplified models improve our understanding of RIS-aided systems since they have tractable expressions yet capture the effect of impedance mismatching and mutual coupling.
These \gls{em}-consistent models can be used for more accurate optimization, resulting in higher performance enabled by the higher channel model accuracy.

\textit{Third}, we further simplify the channel models to get engineering insights into the role of RIS by assuming that all the antennas are perfectly matched and there is no mutual coupling.
We show that these models are consistent with the model widely accepted in related literature, provided that an additional approximation is considered.
We numerically quantify the impact of such an approximation and show that, despite it vanishes as the number of RIS elements increases, it is non-negligible for a practical number of RIS elements.

\textit{Fourth}, we characterize conventional RIS and BD-RIS architectures by using $Z$-, $Y$-, and $S$-parameters.
Since the three descriptions are equivalent, we discuss how to select the most suitable parameters to describe the RIS architecture according to the considered scenario, enabling an effective mathematical characterization of RIS.
While $Z$- and $S$-parameters have been widely used to characterize conventional RIS architectures with and without mutual coupling, respectively, we show that the $Y$-parameters are particularly useful for BD-RIS.
In addition, we illustrate the advantages of the three descriptions in optimizing conventional RIS and BD-RIS architectures by studying three optimization problems.

\textit{Organization}:
In Section~\ref{sec:analysis}, we introduce multiport network analysis.
In Section~\ref{sec:model-general}, we derive general RIS-aided communication models based on $Z$-, $Y$-, and $S$-parameters.
In Section~\ref{sec:model-simplified1}, we simplify the derived models by using the unilateral approximation.
In Section~\ref{sec:model-simplified2}, we further simplify the models assuming perfect matching and no mutual coupling, and compare them with the widely used RIS-aided channel model in communication engineering.
In Section~\ref{sec:architecture-optimization}, we discuss the advantages of using the different parameters in characterizing and optimizing an RIS.
In Section~\ref{sec:results}, we evaluate the performance of an RIS-aided system obtained by solving the presented case studies.
Finally, Section~\ref{sec:conclusion} concludes this work.

\textit{Notation}:
Vectors and matrices are denoted with bold lower and bold upper letters, respectively.
Scalars are represented with letters not in bold font.
$\vert a\vert$ and arg$(a)$ refer to the absolute value and phase of a complex scalar $a$.
$[\mathbf{a}]_{i}$ and $\Vert\mathbf{a}\Vert$ refer to the $i$th element and $l_{2}$-norm of a vector $\mathbf{a}$, respectively.
$\mathbf{A}^T$, $\mathbf{A}^H$, $[\mathbf{A}]_{i,j}$, and $\Vert\mathbf{A}\Vert_F$ refer to the transpose, conjugate transpose, $(i,j)$th element, and Frobenius norm of a matrix $\mathbf{A}$, respectively.
$\mathbb{R}$ and $\mathbb{C}$ denote real and complex number sets, respectively.
$j=\sqrt{-1}$ denotes the imaginary unit.
$\mathbf{0}$ and $\mathbf{I}$ denote an all-zero matrix and an identity matrix with appropriate dimensions, respectively.
$\mathcal{CN}(\mathbf{0},\mathbf{I})$ denotes the distribution of a circularly symmetric complex Gaussian random vector with mean vector $\mathbf{0}$ and covariance matrix $\mathbf{I}$ and $\sim$ stands for ``distributed as''.
diag$(a_1,\ldots,a_N)$ refers to a diagonal matrix with diagonal elements being $a_1,\ldots,a_N$.
diag$(\mathbf{A}_{1},\ldots,\mathbf{A}_{N})$ refers to a block diagonal matrix with blocks being $\mathbf{A}_{1},\ldots,\mathbf{A}_{N}$.

\begin{figure}[t]
\centering
\includegraphics[width=0.48\textwidth]{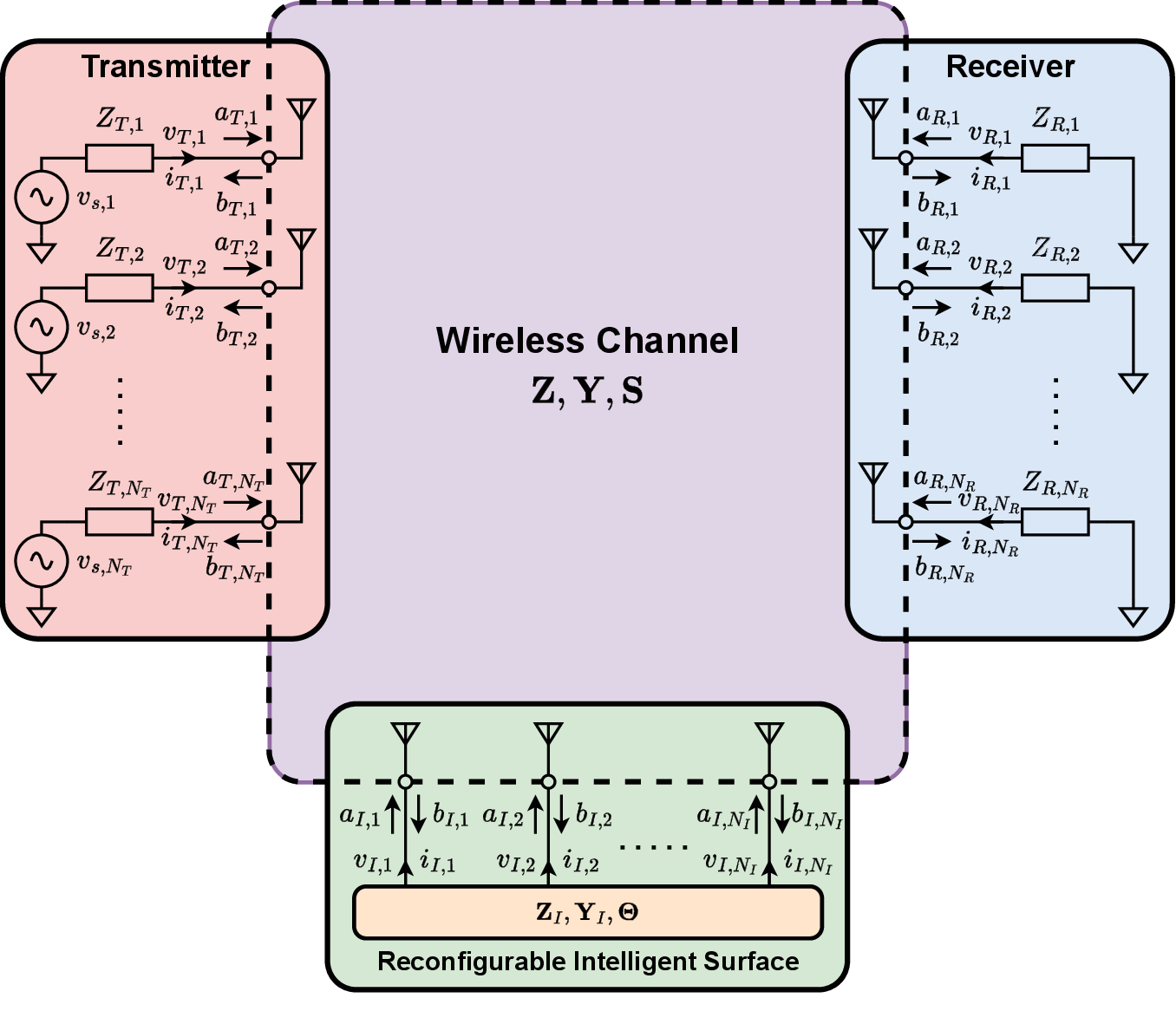}
\caption{Multiport model of an RIS-aided communication system.}
\label{fig:multiport}
\end{figure}

\section{Multiport Network Analysis}
\label{sec:analysis}

We consider a \gls{mimo} communication system aided by an RIS, where there are $N_T$ antennas at the transmitter, $N_R$ antennas at the receiver, and $N_I$ scattering elements at the RIS, as represented in Fig.~\ref{fig:multiport}.
As in \cite{she20}, we model the wireless channel as an $N$-port network, with $N=N_T+N_I+N_R$.
According to multiport network analysis \cite[Chapter 4]{poz11}, this $N$-port network can be characterized by using its impedance, admittance, or scattering matrix, as described in the following\footnote{Note that the considered multiport model can represent a single-user \gls{mimo} system between an $N_T$-antenna transmitter and an $N_R$-antenna receiver, as well as a multi-user \gls{mimo} system with $K$ receivers where the $k$th receiver has $N_k$ antennas, with $\sum_{k=1}^KN_k=N_R$ (see \cite{mur23} for an example).}.

\subsection{Modeling Based on the Impedance Parameters}

The $N$-port network model for a wireless channel can be characterized by its impedance matrix $\mathbf{Z}\in\mathbb{C}^{N\times N}$, such that
\begin{equation}
\mathbf{v}=\mathbf{Z}\mathbf{i},\label{eq:vZi}
\end{equation}
where $\mathbf{v}\in\mathbb{C}^{N\times 1}$ and $\mathbf{i}\in\mathbb{C}^{N\times 1}$ are the voltages and currents at the $N$ ports, respectively.
We can partition $\mathbf{v}$, $\mathbf{i}$, and $\mathbf{Z}$ as
\begin{equation}
\mathbf{v}=
\begin{bmatrix}
\mathbf{v}_{T}\\
\mathbf{v}_{I}\\
\mathbf{v}_{R}
\end{bmatrix},\;
\mathbf{i}=
\begin{bmatrix}
\mathbf{i}_{T}\\
\mathbf{i}_{I}\\
\mathbf{i}_{R}
\end{bmatrix},\;
\mathbf{Z}=
\begin{bmatrix}
\mathbf{Z}_{TT} & \mathbf{Z}_{TI} & \mathbf{Z}_{TR}\\
\mathbf{Z}_{IT} & \mathbf{Z}_{II} & \mathbf{Z}_{IR}\\
\mathbf{Z}_{RT} & \mathbf{Z}_{RI} & \mathbf{Z}_{RR}
\end{bmatrix},
\end{equation}
where $\mathbf{v}_{i}\in\mathbb{C}^{N_i\times 1}$ and $\mathbf{i}_{i}\in\mathbb{C}^{N_i\times 1}$ for $i\in\{T,I,R\}$ refer to the voltages and currents at the antennas of the transmitter, RIS, and receiver, respectively.
Accordingly, $\mathbf{Z}_{TT}\in\mathbb{C}^{N_T\times N_T}$, $\mathbf{Z}_{II}\in\mathbb{C}^{N_I\times N_I}$, and $\mathbf{Z}_{RR}\in\mathbb{C}^{N_R\times N_R}$ refer to the impedance matrices of the antenna arrays at the transmitter, RIS, and receiver, respectively.
The diagonal entries of $\mathbf{Z}_{TT}$, $\mathbf{Z}_{II}$, and $\mathbf{Z}_{RR}$ refer to the antenna self-impedance while the off-diagonal entries refer to antenna mutual coupling.
$\mathbf{Z}_{RT}\in\mathbb{C}^{N_R\times N_T}$, $\mathbf{Z}_{IT}\in\mathbb{C}^{N_I\times N_T}$, and $\mathbf{Z}_{RI}\in\mathbb{C}^{N_R\times N_I}$ refer to the transmission impedance matrices from the transmitter to receiver, from the transmitter to RIS, and from the RIS to receiver, respectively.
Similarly, $\mathbf{Z}_{TR}\in\mathbb{C}^{N_T\times N_R}$, $\mathbf{Z}_{TI}\in\mathbb{C}^{N_T\times N_I}$, and $\mathbf{Z}_{IR}\in\mathbb{C}^{N_I\times N_R}$ refer to the transmission impedance matrices from the receiver to transmitter, from the RIS to transmitter, and from the receiver to RIS, respectively.
Thus, in the case of a reciprocal wireless channel, we have $\mathbf{Z}_{TR}=\mathbf{Z}_{RT}^T$, $\mathbf{Z}_{TI}=\mathbf{Z}_{IT}^T$, and $\mathbf{Z}_{IR}=\mathbf{Z}_{RI}^T$.

At the transmitter, the $n_T$th antenna is connected in series with a source voltage $v_{s,n_T}$ and a source impedance $Z_{T,n_T}$, for $n_T=1,\ldots,N_T$.
Therefore, $\mathbf{v}_T$ and $\mathbf{i}_T$ are related by
\begin{equation}
\mathbf{v}_T=\mathbf{v}_{s,T}-\mathbf{Z}_T\mathbf{i}_T,\label{eq:TX-z}
\end{equation}
where $\mathbf{v}_{s,T}=[v_{s,1},v_{s,2},\ldots,v_{s,N_T}]^T\in\mathbb{C}^{N_T\times 1}$ refers to the source voltage vector and $\mathbf{Z}_{T}\in\mathbb{C}^{N_T\times N_T}$ is a diagonal matrix given by $\mathbf{Z}_{T}=\textrm{diag}(Z_{T,1},Z_{T,2},\ldots,Z_{T,N_T})$.
At the RIS, the $N_I$ scattering elements are connected to an $N_I$-port reconfigurable impedance network and $\mathbf{v}_I$ is related to $\mathbf{i}_I$ by
\begin{equation}
\mathbf{v}_I=-\mathbf{Z}_I\mathbf{i}_I,\label{eq:RIS-z}
\end{equation}
where $\mathbf{Z}_{I}\in\mathbb{C}^{N_I\times N_I}$ is the impedance matrix of the $N_I$-port reconfigurable impedance network.
At the receiver, the $n_R$th antenna is connected in series with a load impedance $Z_{R,n_R}$, for $n_R=1,\ldots,N_R$.
Therefore, $\mathbf{v}_R$ and $\mathbf{i}_R$ are related by
\begin{equation}
\mathbf{v}_R=-\mathbf{Z}_R\mathbf{i}_R,\label{eq:RX-z}
\end{equation}
where $\mathbf{Z}_{R}\in\mathbb{C}^{N_R\times N_R}$ is a diagonal matrix given by $\mathbf{Z}_{R}=\textrm{diag}(Z_{R,1},Z_{R,2},\ldots,Z_{R,N_R})$.

\subsection{Modeling Based on the Admittance Parameters}

The wireless channel can also be characterized by its admittance matrix $\mathbf{Y}\in\mathbb{C}^{N\times N}$, with $\mathbf{Y}=\mathbf{Z}^{-1}$, such that
\begin{equation}
\mathbf{i}=\mathbf{Y}\mathbf{v}.\label{eq:iYv}
\end{equation}
Similar to $\mathbf{Z}$, also $\mathbf{Y}$ can be partitioned as
\begin{equation}
\mathbf{Y}=
\begin{bmatrix}
\mathbf{Y}_{TT} & \mathbf{Y}_{TI} & \mathbf{Y}_{TR}\\
\mathbf{Y}_{IT} & \mathbf{Y}_{II} & \mathbf{Y}_{IR}\\
\mathbf{Y}_{RT} & \mathbf{Y}_{RI} & \mathbf{Y}_{RR}
\end{bmatrix},
\end{equation}
where $\mathbf{Y}_{ij}\in\mathbb{C}^{N_i\times N_j}$ for $i,j\in\{T,I,R\}$.

At the transmitter, $\mathbf{v}_T$ and $\mathbf{i}_T$ are related by
\begin{equation}
\mathbf{i}_T=\mathbf{i}_{s,T}-\mathbf{Y}_T\mathbf{v}_T,\label{eq:TX-y}
\end{equation}
where $\mathbf{i}_{s,T}\in\mathbb{C}^{N_T\times 1}$ refers to the source current vector and $\mathbf{Y}_{T}\in\mathbb{C}^{N_T\times N_T}$ is a diagonal matrix given by $\mathbf{Y}_T=\mathbf{Z}_T^{-1}$.
At the RIS, $\mathbf{v}_I$ and $\mathbf{i}_I$ are related by
\begin{equation}
\mathbf{i}_I=-\mathbf{Y}_I\mathbf{v}_I,\label{eq:RIS-y}
\end{equation}
where $\mathbf{Y}_{I}\in\mathbb{C}^{N_I\times N_I}$ is the admittance matrix of the $N_I$-port reconfigurable impedance network, given by $\mathbf{Y}_I=\mathbf{Z}_I^{-1}$.
At the receiver, $\mathbf{v}_R$ and $\mathbf{i}_R$ are related by
\begin{equation}
\mathbf{i}_R=-\mathbf{Y}_R\mathbf{v}_R,\label{eq:RX-y}
\end{equation}
where $\mathbf{Y}_{R}\in\mathbb{C}^{N_R\times N_R}$ is diagonal given by $\mathbf{Y}_R=\mathbf{Z}_R^{-1}$.

\subsection{Modeling Based on the Scattering Parameters}

Finally, the wireless channel can be characterized by its scattering matrix $\mathbf{S}\in\mathbb{C}^{N\times N}$, so that we have
\begin{equation}
\mathbf{b}=\mathbf{S}\mathbf{a},\label{eq:bSa}
\end{equation}
where $\mathbf{a}\in\mathbb{C}^{N\times 1}$ and $\mathbf{b}\in\mathbb{C}^{N\times 1}$ are the incident and reflected waves at the ports, respectively.
We partition $\mathbf{a}$, $\mathbf{b}$, and $\mathbf{S}$ as
\begin{equation}
\mathbf{a}=
\begin{bmatrix}
\mathbf{a}_{T}\\
\mathbf{a}_{I}\\
\mathbf{a}_{R}
\end{bmatrix},\;
\mathbf{b}=
\begin{bmatrix}
\mathbf{b}_{T}\\
\mathbf{b}_{I}\\
\mathbf{b}_{R}
\end{bmatrix},\;
\mathbf{S}=
\begin{bmatrix}
\mathbf{S}_{TT} & \mathbf{S}_{TI} & \mathbf{S}_{TR}\\
\mathbf{S}_{IT} & \mathbf{S}_{II} & \mathbf{S}_{IR}\\
\mathbf{S}_{RT} & \mathbf{S}_{RI} & \mathbf{S}_{RR}
\end{bmatrix},
\end{equation}
where $\mathbf{a}_{i}\in\mathbb{C}^{N_i\times 1}$ and $\mathbf{b}_{i}\in\mathbb{C}^{N_i\times 1}$ for $i\in\{T,I,R\}$ refer to the incident and reflected waves at the antennas of the transmitter, RIS, and receiver, respectively, and $\mathbf{S}_{ij}\in\mathbb{C}^{N_i\times N_j}$ for $i,j\in\{T,I,R\}$.
Remarkably, according to \cite[Chapter 4]{poz11}, the vectors $\mathbf{v}$ and $\mathbf{i}$ are related to $\mathbf{a}$ and $\mathbf{b}$ though
\begin{equation}
\mathbf{v}=\mathbf{a}+\mathbf{b},\;\mathbf{i}=\frac{\mathbf{a}-\mathbf{b}}{Z_0}=Y_0\left(\mathbf{a}-\mathbf{b}\right),\label{eq:viab}
\end{equation}
respectively, where $Z_0$ is the characteristic impedance used to compute the $S$-parameters, typically equal to $Z_0=50$~$\Omega$, and $Y_0=1/Z_0$ is the characteristic admittance.
In addition, the matrix $\mathbf{S}$ can be expressed as a function of $\mathbf{Z}$ as
\begin{equation}
\mathbf{S}=\left(\mathbf{Z}+Z_0\mathbf{I}\right)^{-1}\left(\mathbf{Z}-Z_0\mathbf{I}\right).\label{eq:S}
\end{equation}

At the transmitter, $\mathbf{a}_T$ and $\mathbf{b}_T$ are related by
\begin{equation}
\mathbf{a}_T=\mathbf{b}_{s,T}+\mathbf{\Gamma}_T\mathbf{b}_T,\label{eq:TX-s}
\end{equation}
where $\mathbf{b}_{s,T}\in\mathbb{C}^{N_T\times 1}$ refers to the source wave vector and $\mathbf{\Gamma}_{T}\in\mathbb{C}^{N_T\times N_T}$ is diagonal with its $(n_T,n_T)$th element being the reflection coefficient of the $n_T$th source impedance, i.e.,
\begin{equation}
\mathbf{\Gamma}_T=\left(\mathbf{Z}_T+Z_0\mathbf{I}\right)^{-1}\left(\mathbf{Z}_T-Z_0\mathbf{I}\right).\label{eq:GammaT}
\end{equation}
At the RIS, $\mathbf{a}_I$ and $\mathbf{b}_I$ are related by
\begin{equation}
\mathbf{a}_I=\boldsymbol{\Theta}\mathbf{b}_I,\label{eq:RIS-s}
\end{equation}
where $\boldsymbol{\Theta}\in\mathbb{C}^{N_I\times N_I}$ denotes the scattering matrix of the $N_I$-port reconfigurable impedance network, written as
\begin{equation}
\boldsymbol{\Theta}=\left(\mathbf{Z}_I+Z_0\mathbf{I}\right)^{-1}\left(\mathbf{Z}_I-Z_0\mathbf{I}\right).\label{eq:Theta}
\end{equation}
At the receiver, $\mathbf{a}_R$ and $\mathbf{b}_R$ are related by
\begin{equation}
\mathbf{a}_R=\mathbf{\Gamma}_R\mathbf{b}_R,\label{eq:RX-s}
\end{equation}
where $\mathbf{\Gamma}_{R}\in\mathbb{C}^{N_R\times N_R}$ is diagonal with its $(n_R,n_R)$th element being the reflection coefficient of the $n_R$th load, i.e.,
\begin{equation}
\mathbf{\Gamma}_R=\left(\mathbf{Z}_R+Z_0\mathbf{I}\right)^{-1}\left(\mathbf{Z}_R-Z_0\mathbf{I}\right).\label{eq:GammaR}
\end{equation}

\section{General RIS-Aided Communication Model}
\label{sec:model-general}

We have characterized the relationships between the electrical properties at the transmitter, RIS, and receiver of an RIS-aided communication system.
In this section, we determine the channel $\mathbf{H}\in\mathbb{C}^{N_R\times N_T}$ relating the voltage vector at the transmitter (transmitted signal) $\mathbf{v}_T$ and the voltage vector at the receiver (received signal) $\mathbf{v}_R$ through $\mathbf{v}_R=\mathbf{H}\mathbf{v}_T$.
To this end, we introduce a universal framework enabling three equivalent analyses, based on $Z$-, $Y$-, and $S$-parameters\footnote{Note that our definition of $\mathbf{H}$ is the one commonly adopted in wireless communications, even though other definitions have been used, such as $\mathbf{v}_R=\mathbf{H}\mathbf{v}_{s,T}$ in \cite{gra21} and $\mathbf{b}_R=\mathbf{H}\mathbf{v}_{s,T}$ in \cite{abr23}.}.

\subsection{Universal Framework}
\label{sec:general-universal}

Consider the following mathematical problem.
Given two variable vectors $\mathbf{x}\in\mathbb{C}^{N\times 1}$ and $\mathbf{y}\in\mathbb{C}^{N\times 1}$, and a constant matrix $\mathbf{A}\in\mathbb{C}^{N\times N}$, respectively partitioned as
\begin{equation}
\mathbf{x}=
\begin{bmatrix}
\mathbf{x}_{1}\\
\mathbf{x}_{2}\\
\mathbf{x}_{3}
\end{bmatrix},\;
\mathbf{y}=
\begin{bmatrix}
\mathbf{y}_{1}\\
\mathbf{y}_{2}\\
\mathbf{y}_{3}
\end{bmatrix},\;
\mathbf{A}=
\begin{bmatrix}
\mathbf{A}_{11} & \mathbf{A}_{12} & \mathbf{A}_{13}\\
\mathbf{A}_{21} & \mathbf{A}_{22} & \mathbf{A}_{23}\\
\mathbf{A}_{31} & \mathbf{A}_{32} & \mathbf{A}_{33}
\end{bmatrix},
\end{equation}
where $\mathbf{x}_{i}\in\mathbb{C}^{N_i\times 1}$, $\mathbf{y}_{i}\in\mathbb{C}^{N_i\times 1}$, $\mathbf{A}_{ij}\in\mathbb{C}^{N_i\times N_j}$ for $i,j\in\{1,2,3\}$, and $N_1+N_2+N_3=N$, our goal is to solve
\begin{equation}
\begin{cases}
\mathbf{y}=\mathbf{A}\mathbf{x},\\
\mathbf{x}_1=\mathbf{c}_1+\mathbf{A}_1\mathbf{y}_1,\\
\mathbf{x}_2=\mathbf{A}_2\mathbf{y}_2,\\
\mathbf{x}_3=\mathbf{A}_3\mathbf{y}_3,
\end{cases}\label{eq:sys-gen1}
\end{equation}
where $\mathbf{c}_1\in\mathbb{C}^{N_1\times 1}$ and $\mathbf{A}_i\in\mathbb{C}^{N_i\times N_i}$ for $i\in\{1,2,3\}$ are constant.
In other words, we want to characterize the two variable vectors $\mathbf{x}$ and $\mathbf{y}$ as a function of the constants $\mathbf{A}$, $\mathbf{c}_1$, $\mathbf{A}_1$, $\mathbf{A}_2$, and $\mathbf{A}_3$.
Remarkably, system \eqref{eq:sys-gen1} provides a universal framework that can be used to describe the equations of the multiport network analysis based on $Z$-, $Y$-, and $S$-parameters.
In Tab.~\ref{tab:universal-mapping}, we report the relationship between the variables in this framework and the quantities in the three different parameters introduced in Section~\ref{sec:analysis}.

\begin{table}[t]
\centering
\caption{Relationship between the universal framework\\and the $Z$-, $Y$-, and $S$-parameters.}
\begin{tabular}{@{}cccc@{}}
\toprule
Framework & $Z$-parameters & $Y$-parameters & $S$-parameters\\
\midrule
$\mathbf{x}$ & $\mathbf{v}$ & $\mathbf{i}$ & $\mathbf{a}$\\
\midrule
$\mathbf{y}$ & $\mathbf{i}$ & $\mathbf{v}$ & $\mathbf{b}$\\
\midrule
$\mathbf{A}$ & $\mathbf{Z}^{-1}$ & $\mathbf{Y}^{-1}$ & $\mathbf{S}$\\
\midrule
$\mathbf{c}_{1}$ & $\mathbf{v}_{s,T}$ & $\mathbf{i}_{s,T}$ & $\mathbf{b}_{s,T}$\\
\midrule
$\mathbf{A}_{1}$ & $-\mathbf{Z}_{T}$ & $-\mathbf{Y}_{T}$ & $\mathbf{\Gamma}_{T}$\\
\midrule
$\mathbf{A}_{2}$ & $-\mathbf{Z}_{I}$ & $-\mathbf{Y}_{I}$ & $\boldsymbol{\Theta}$\\
\midrule
$\mathbf{A}_{3}$ & $-\mathbf{Z}_{R}$ & $-\mathbf{Y}_{R}$ & $\mathbf{\Gamma}_{R}$\\
\bottomrule
\end{tabular}
\label{tab:universal-mapping}
\end{table}

To solve system \eqref{eq:sys-gen1}, we rewrite it in a compact form
\begin{equation}
\begin{cases}
\mathbf{y}=\mathbf{A}\mathbf{x},\\
\mathbf{x}=\mathbf{c}+\overline{\mathbf{A}}\mathbf{y},
\end{cases}\label{eq:sys-gen2}
\end{equation}
where we introduce $\mathbf{c}\in\mathbb{C}^{N\times 1}$ and $\overline{\mathbf{A}}\in\mathbb{C}^{N\times N}$ as
\begin{equation}
\mathbf{c}=
\begin{bmatrix}
\mathbf{c}_1\\
\mathbf{0}\\
\mathbf{0}
\end{bmatrix},\;
\overline{\mathbf{A}}=
\begin{bmatrix}
\mathbf{A}_{1} & \mathbf{0} & \mathbf{0}\\
\mathbf{0} & \mathbf{A}_{2} & \mathbf{0}\\
\mathbf{0} & \mathbf{0} & \mathbf{A}_{3}
\end{bmatrix}.
\end{equation}
System \eqref{eq:sys-gen2} can be solved by noting that
\begin{equation}
\mathbf{x}=\mathbf{c}+\overline{\mathbf{A}}\mathbf{A}\mathbf{x},
\end{equation}
gives
\begin{equation}
\mathbf{x}=\left(\mathbf{I}-\overline{\mathbf{A}}\mathbf{A}\right)^{-1}\mathbf{c}.\label{eq:sol-gen}
\end{equation}
Thus, by introducing $\widetilde{\mathbf{A}}=\left(\mathbf{I}-\overline{\mathbf{A}}\mathbf{A}\right)^{-1}$, partitioned as
\begin{equation}
\widetilde{\mathbf{A}}=
\begin{bmatrix}
\widetilde{\mathbf{A}}_{11} & \widetilde{\mathbf{A}}_{12} & \widetilde{\mathbf{A}}_{13}\\
\widetilde{\mathbf{A}}_{21} & \widetilde{\mathbf{A}}_{22} & \widetilde{\mathbf{A}}_{23}\\
\widetilde{\mathbf{A}}_{31} & \widetilde{\mathbf{A}}_{32} & \widetilde{\mathbf{A}}_{33}
\end{bmatrix},
\end{equation}
we have
\begin{equation}
\mathbf{x}_1=\widetilde{\mathbf{A}}_{11}\mathbf{c}_1,\;\mathbf{x}_2=\widetilde{\mathbf{A}}_{21}\mathbf{c}_1,\;\mathbf{x}_3=\widetilde{\mathbf{A}}_{31}\mathbf{c}_1,\label{eq:sol-univ}
\end{equation}
providing an expression for the vector $\mathbf{x}$.
Given $\mathbf{x}$, the vector $\mathbf{y}$ is determined by applying $\mathbf{y}=\mathbf{A}\mathbf{x}$ and system \eqref{eq:sys-gen1} is solved.

\subsection{Impedance Parameters Analysis}

With the $Z$-parameters, the channel $\mathbf{H}$ is derived by solving the system of the four linear equations \eqref{eq:vZi}, \eqref{eq:TX-z}, \eqref{eq:RIS-z}, and \eqref{eq:RX-z}.
According to Sec.~\ref{sec:general-universal}, this system gives
\begin{equation}
\mathbf{v}=\left(\mathbf{I}+\overline{\mathbf{Z}}\mathbf{Z}^{-1}\right)^{-1}\mathbf{v}_s,\label{eq:sol-z1}
\end{equation}
where we define $\mathbf{v}_s\in\mathbb{C}^{N\times 1}$ and $\overline{\mathbf{Z}}\in\mathbb{C}^{N\times N}$ as
\begin{equation}
\mathbf{v}_s=
\begin{bmatrix}
\mathbf{v}_{s,T}\\
\mathbf{0}\\
\mathbf{0}
\end{bmatrix},\;
\overline{\mathbf{Z}}=
\begin{bmatrix}
\mathbf{Z}_{T} & \mathbf{0} & \mathbf{0}\\
\mathbf{0} & \mathbf{Z}_{I} & \mathbf{0}\\
\mathbf{0} & \mathbf{0} & \mathbf{Z}_{R}
\end{bmatrix}.
\end{equation}
Thus, by introducing $\widetilde{\mathbf{Z}}=\left(\mathbf{I}+\overline{\mathbf{Z}}\mathbf{Z}^{-1}\right)^{-1}$, partitioned as
\begin{equation}
\widetilde{\mathbf{Z}}=
\begin{bmatrix}
\widetilde{\mathbf{Z}}_{TT} & \widetilde{\mathbf{Z}}_{TI} & \widetilde{\mathbf{Z}}_{TR}\\
\widetilde{\mathbf{Z}}_{IT} & \widetilde{\mathbf{Z}}_{II} & \widetilde{\mathbf{Z}}_{IR}\\
\widetilde{\mathbf{Z}}_{RT} & \widetilde{\mathbf{Z}}_{RI} & \widetilde{\mathbf{Z}}_{RR}
\end{bmatrix},
\end{equation}
we have
\begin{equation}
\mathbf{v}_T=\widetilde{\mathbf{Z}}_{TT}\mathbf{v}_{s,T},\;\mathbf{v}_R=\widetilde{\mathbf{Z}}_{RT}\mathbf{v}_{s,T}.\label{eq:sol-z2}
\end{equation}
As a consequence of \eqref{eq:sol-z2}, we obtain
\begin{equation}
\mathbf{v}_R=\widetilde{\mathbf{Z}}_{RT}\mathbf{v}_{s,T}=\widetilde{\mathbf{Z}}_{RT}\widetilde{\mathbf{Z}}_{TT}^{-1}\mathbf{v}_{T},
\end{equation}
yielding
\begin{equation}
\mathbf{H}=\widetilde{\mathbf{Z}}_{RT}\widetilde{\mathbf{Z}}_{TT}^{-1},\label{eq:Hgen-z}
\end{equation}
which is the general channel model based on the $Z$-parameters.

\subsection{Admittance Parameters Analysis}

With the $Y$-parameters, the channel $\mathbf{H}$ is derived by solving the system of the four linear equations \eqref{eq:iYv}, \eqref{eq:TX-y}, \eqref{eq:RIS-y}, and \eqref{eq:RX-y}.
According to Sec.~\ref{sec:general-universal}, this system gives
\begin{equation}
\mathbf{i}=\left(\mathbf{I}+\overline{\mathbf{Y}}\mathbf{Y}^{-1}\right)^{-1}\mathbf{i}_s,\label{eq:sol-y1}
\end{equation}
where we define $\mathbf{i}_s\in\mathbb{C}^{N\times 1}$ and $\overline{\mathbf{Y}}\in\mathbb{C}^{N\times N}$ as
\begin{equation}
\mathbf{i}_s=
\begin{bmatrix}
\mathbf{i}_{s,T}\\
\mathbf{0}\\
\mathbf{0}
\end{bmatrix},\;
\overline{\mathbf{Y}}=
\begin{bmatrix}
\mathbf{Y}_{T} & \mathbf{0} & \mathbf{0}\\
\mathbf{0} & \mathbf{Y}_{I} & \mathbf{0}\\
\mathbf{0} & \mathbf{0} & \mathbf{Y}_{R}
\end{bmatrix}.
\end{equation}
Thus, by introducing $\widetilde{\mathbf{Y}}=\left(\mathbf{I}+\overline{\mathbf{Y}}\mathbf{Y}^{-1}\right)^{-1}$, partitioned as
\begin{equation}
\widetilde{\mathbf{Y}}=
\begin{bmatrix}
\widetilde{\mathbf{Y}}_{TT} & \widetilde{\mathbf{Y}}_{TI} & \widetilde{\mathbf{Y}}_{TR}\\
\widetilde{\mathbf{Y}}_{IT} & \widetilde{\mathbf{Y}}_{II} & \widetilde{\mathbf{Y}}_{IR}\\
\widetilde{\mathbf{Y}}_{RT} & \widetilde{\mathbf{Y}}_{RI} & \widetilde{\mathbf{Y}}_{RR}
\end{bmatrix},
\end{equation}
we have
\begin{equation}
\mathbf{i}_T=\widetilde{\mathbf{Y}}_{TT}\mathbf{i}_{s,T},\;\mathbf{i}_R=\widetilde{\mathbf{Y}}_{RT}\mathbf{i}_{s,T}.\label{eq:sol-y2}
\end{equation}
As a consequence of \eqref{eq:TX-y}, \eqref{eq:RX-y}, and \eqref{eq:sol-y2}, we obtain
\begin{align}
\mathbf{v}_R
&=-\mathbf{Y}_{R}^{-1}\mathbf{i}_R=-\mathbf{Y}_{R}^{-1}\widetilde{\mathbf{Y}}_{RT}\mathbf{i}_{s,T}\\
&=\mathbf{Y}_{R}^{-1}\widetilde{\mathbf{Y}}_{RT}\left(\widetilde{\mathbf{Y}}_{TT}-\mathbf{I}\right)^{-1}\mathbf{Y}_{T}\mathbf{v}_{T},
\end{align}
yielding
\begin{equation}
\mathbf{H}=\mathbf{Y}_{R}^{-1}\widetilde{\mathbf{Y}}_{RT}\left(\widetilde{\mathbf{Y}}_{TT}-\mathbf{I}\right)^{-1}\mathbf{Y}_{T},\label{eq:Hgen-y}
\end{equation}
which is the general channel model with the $Y$-parameters.

\subsection{Scattering Parameters Analysis}

With the $S$-parameters, the channel $\mathbf{H}$ is derived by solving the system of \eqref{eq:bSa}, \eqref{eq:TX-s}, \eqref{eq:RIS-s}, and \eqref{eq:RX-s}.
According to Sec.~\ref{sec:general-universal}, and using $\mathbf{b}=\mathbf{S}\mathbf{a}$, this system gives
\begin{equation}
\mathbf{b}=\mathbf{S}\left(\mathbf{I}-\mathbf{\Gamma}\mathbf{S}\right)^{-1}\mathbf{b}_{s},\label{eq:sol-s1}
\end{equation}
where we define $\mathbf{b}_s\in\mathbb{C}^{N\times 1}$ and $\mathbf{\Gamma}\in\mathbb{C}^{N\times N}$ as
\begin{equation}
\mathbf{b}_s=
\begin{bmatrix}
\mathbf{b}_{s,T}\\
\mathbf{0}\\
\mathbf{0}
\end{bmatrix},\;
\mathbf{\Gamma}=
\begin{bmatrix}
\mathbf{\Gamma}_{T} & \mathbf{0} & \mathbf{0}\\
\mathbf{0} & \boldsymbol{\Theta} & \mathbf{0}\\
\mathbf{0} & \mathbf{0} & \mathbf{\Gamma}_{R}
\end{bmatrix}.
\end{equation}
Thus, by introducing $\widetilde{\mathbf{S}}=\mathbf{S}\left(\mathbf{I}-\mathbf{\Gamma}\mathbf{S}\right)^{-1}$, partitioned as
\begin{equation}
\widetilde{\mathbf{S}}=
\begin{bmatrix}
\widetilde{\mathbf{S}}_{TT} & \widetilde{\mathbf{S}}_{TI} & \widetilde{\mathbf{S}}_{TR}\\
\widetilde{\mathbf{S}}_{IT} & \widetilde{\mathbf{S}}_{II} & \widetilde{\mathbf{S}}_{IR}\\
\widetilde{\mathbf{S}}_{RT} & \widetilde{\mathbf{S}}_{RI} & \widetilde{\mathbf{S}}_{RR}
\end{bmatrix},
\end{equation}
we have
\begin{equation}
\mathbf{b}_T=\widetilde{\mathbf{S}}_{TT}\mathbf{b}_{s,T},\;\mathbf{b}_R=\widetilde{\mathbf{S}}_{RT}\mathbf{b}_{s,T}.\label{eq:sol-s2}
\end{equation}
As a consequence of \eqref{eq:viab}, \eqref{eq:TX-s}, \eqref{eq:RX-s}, and \eqref{eq:sol-s2}, we obtain
\begin{align}
\mathbf{v}_R
&=\mathbf{a}_R+\mathbf{b}_R=\left(\mathbf{\Gamma}_{R}+\mathbf{I}\right)\widetilde{\mathbf{S}}_{RT}\mathbf{b}_{s,T}\\
&=\left(\mathbf{\Gamma}_{R}+\mathbf{I}\right)\widetilde{\mathbf{S}}_{RT}\left(\mathbf{I}+\mathbf{\Gamma}_{T}\widetilde{\mathbf{S}}_{TT}+\widetilde{\mathbf{S}}_{TT}\right)^{-1}\mathbf{v}_{T},
\end{align}
yielding
\begin{equation}
\mathbf{H}=\left(\mathbf{\Gamma}_{R}+\mathbf{I}\right)\widetilde{\mathbf{S}}_{RT}\left(\mathbf{I}+\mathbf{\Gamma}_{T}\widetilde{\mathbf{S}}_{TT}+\widetilde{\mathbf{S}}_{TT}\right)^{-1},\label{eq:Hgen-s}
\end{equation}
which is the general channel model based on the $S$-parameters, in agreement with \cite{she20}.

\subsection{Equivalence Between General Models}

We have derived three general channel models in \eqref{eq:Hgen-z}, \eqref{eq:Hgen-y}, and \eqref{eq:Hgen-s}, based on $Z$-, $Y$-, and $S$-parameters, respectively.
We now confirm, for the first time, that these channel models are equivalent representations of an RIS-aided system.

For this purpose, it is necessary to relate the vectors $\mathbf{i}_{s,T}$ and $\mathbf{b}_{s,T}$ to the vector $\mathbf{v}_{s,T}$ by observing that \eqref{eq:TX-z}, \eqref{eq:TX-y}, and \eqref{eq:TX-s} are three equivalent descriptions of the electrical properties at the transmitter.
By equating \eqref{eq:TX-z} and \eqref{eq:TX-y}, and recalling that $\mathbf{Y}_T=\mathbf{Z}_T^{-1}$, we obtain
\begin{equation}
\mathbf{i}_{s,T}=\mathbf{Y}_T\mathbf{v}_{s,T},\label{eq:is}
\end{equation}
giving the relationship between $\mathbf{i}_{s,T}$ and $\mathbf{v}_{s,T}$.
In addition, by equating \eqref{eq:TX-z} and \eqref{eq:TX-s}, and using \eqref{eq:viab} and \eqref{eq:GammaT}, we obtain
\begin{equation}
\mathbf{b}_{s,T}=\frac{\mathbf{I}-\mathbf{\Gamma}_T}{2}\mathbf{v}_{s,T},\label{eq:bs}
\end{equation}
giving the relationship between $\mathbf{b}_{s,T}$ and $\mathbf{v}_{s,T}$.
We provide the equivalence of the general channel models in the following two propositions.
\begin{proposition}
The general channel model based on the $Z$-parameters \eqref{eq:Hgen-z} is equivalent to the general channel model based on the $Y$-parameters \eqref{eq:Hgen-y}.
\end{proposition}
\begin{proof}
To show this equivalence, we note that \eqref{eq:Hgen-z} is a direct consequence of \eqref{eq:sol-z1} and that \eqref{eq:Hgen-y} is a direct consequence of \eqref{eq:sol-y1}.
Thus, it is convenient to just show that \eqref{eq:sol-z1} and \eqref{eq:sol-y1} are equivalent.
By using $\overline{\mathbf{Z}}=\overline{\mathbf{Y}}^{-1}$ and $\mathbf{Z}=\mathbf{Y}^{-1}$, we can rewrite \eqref{eq:sol-z1} as
\begin{equation}
\mathbf{v}=\left(\mathbf{I}+\overline{\mathbf{Y}}^{-1}\mathbf{Y}\right)^{-1}\mathbf{v}_s.\label{eq:proof-gen1}
\end{equation}
Then, recalling that $\mathbf{i}=\mathbf{Y}\mathbf{v}$ and observing that \eqref{eq:is} gives $\mathbf{v}_s=\overline{\mathbf{Y}}^{-1}\mathbf{i}_s$, from \eqref{eq:proof-gen1} we obtain
\begin{align}
\mathbf{i}
&=\mathbf{Y}\left(\mathbf{I}+\overline{\mathbf{Y}}^{-1}\mathbf{Y}\right)^{-1}\overline{\mathbf{Y}}^{-1}\mathbf{i}_s\\
&=\left(\mathbf{I}+\overline{\mathbf{Y}}\mathbf{Y}^{-1}\right)^{-1}\mathbf{i}_s,
\end{align}
giving \eqref{eq:sol-y1}.
Thus, we have that the general channel models \eqref{eq:Hgen-z} and \eqref{eq:Hgen-y} are equivalent since they are a direct consequence of equivalent equations.
\end{proof}
\begin{proposition}
The general channel model based on the $Z$-parameters \eqref{eq:Hgen-z} is equivalent to the general channel model based on the $S$-parameters \eqref{eq:Hgen-s}.
\end{proposition}
\begin{proof}
Since \eqref{eq:Hgen-z} is a direct consequence of \eqref{eq:sol-z1} and \eqref{eq:Hgen-s} is a direct consequence of \eqref{eq:sol-s1}, it is convenient to just show that \eqref{eq:sol-z1} and \eqref{eq:sol-s1} are equivalent.
By using $\overline{\mathbf{Z}}=Z_0\left(\mathbf{I}+\mathbf{\Gamma}\right)\left(\mathbf{I}-\mathbf{\Gamma}\right)^{-1}$ and $\mathbf{Z}=Z_0\left(\mathbf{I}+\mathbf{S}\right)\left(\mathbf{I}-\mathbf{S}\right)^{-1}$, we can rewrite \eqref{eq:sol-z1} as
\begin{align}
\mathbf{v}
&=\left(\mathbf{I}+\left(\mathbf{I}+\mathbf{\Gamma}\right)\left(\mathbf{I}-\mathbf{\Gamma}\right)^{-1}\left(\mathbf{I}-\mathbf{S}\right)\left(\mathbf{I}+\mathbf{S}\right)^{-1}\right)^{-1}\mathbf{v}_s\\
&=\left(\mathbf{I}+\mathbf{S}\right)\left(\mathbf{I}-\mathbf{\Gamma}\mathbf{S}\right)^{-1}\left(\mathbf{I}-\mathbf{\Gamma}\right)\mathbf{v}_{s}/2.\label{eq:proof-gen2}
\end{align}
Then, recalling that $\mathbf{b}=(\mathbf{S}^{-1}+\mathbf{I})^{-1}\mathbf{v}$ and observing that \eqref{eq:bs} gives $\mathbf{v}_{s}=2(\mathbf{I}-\mathbf{\Gamma})^{-1}\mathbf{b}_{s}$, from \eqref{eq:proof-gen2} we obtain
\begin{align}
\mathbf{b}
&=\left(\mathbf{S}^{-1}+\mathbf{I}\right)^{-1}\left(\mathbf{I}+\mathbf{S}\right)\left(\mathbf{I}-\mathbf{\Gamma}\mathbf{S}\right)^{-1}\mathbf{b}_{s}\\
&=\mathbf{S}\left(\mathbf{I}-\mathbf{\Gamma}\mathbf{S}\right)^{-1}\mathbf{b}_{s},
\end{align}
giving \eqref{eq:sol-s1}.
Thus, the general channel models \eqref{eq:Hgen-z} and \eqref{eq:Hgen-s} are equivalent since they are a direct consequence of equivalent equations, in agreement with network theory \cite[Chapter 4]{poz11}.
\end{proof}
By the transitive property of the equivalence relation, we also have that the general channel models based on the $Y$- and $S$-parameters are equivalent.

\section{RIS-Aided Communication Model Using the Unilateral Approximation}
\label{sec:model-simplified1}

We have derived three general channel models based on the $Z$-, $Y$-, and $S$-parameters.
They account for imperfect matching and mutual coupling at the transmitter, RIS, and receiver as they have been obtained without any approximation or assumption.
However, it is difficult to obtain insights into the role of the RIS in the communication model due to the presence of matrix inversion operations.
To gain a better understanding of the communication models, in this section, we approximate the general models by assuming sufficiently large transmission distances.

\subsection{Universal Framework}
\label{sec:simplified-universal}

We begin by simplifying the universal framework introduced in Sec.~\ref{sec:general-universal}.
Specifically, we consider the matrix $\mathbf{A}$ to be block lower triangular given by
\begin{equation}
\mathbf{A}=
\begin{bmatrix}
\mathbf{A}_{11} & \mathbf{0} & \mathbf{0}\\
\mathbf{A}_{21} & \mathbf{A}_{22} & \mathbf{0}\\
\mathbf{A}_{31} & \mathbf{A}_{32} & \mathbf{A}_{33}
\end{bmatrix}.\label{eq:A-uni}
\end{equation}
Remarkably, system \eqref{eq:sys-gen1} with $\mathbf{A}$ given by \eqref{eq:A-uni} provides a universal framework that can be used to describe the equations based on $Z$-, $Y$-, and $S$-parameters with the unilateral approximation \cite{ivr10}, as it will be clarified in Sec.~\ref{sec:simplified-z}, \ref{sec:simplified-y}, and \ref{sec:simplified-s}, respectively.
With the simplification in \eqref{eq:A-uni}, the solution to system \eqref{eq:sys-gen1} given by \eqref{eq:sol-univ} simplifies as
\begin{equation}
\mathbf{x}_1=\left(\mathbf{I}-\mathbf{A}_{1}\mathbf{A}_{11}\right)^{-1}\mathbf{c}_1,
\end{equation}
\begin{equation}
\mathbf{x}_2=\left(\mathbf{I}-\mathbf{A}_{2}\mathbf{A}_{22}\right)^{-1}\mathbf{A}_{2}\mathbf{A}_{21}\left(\mathbf{I}-\mathbf{A}_{1}\mathbf{A}_{11}\right)^{-1}\mathbf{c}_1,
\end{equation}
\begin{multline}
\mathbf{x}_3=\left(\mathbf{A}_{3}^{-1}-\mathbf{A}_{33}\right)^{-1}\left(\mathbf{A}_{31}+\mathbf{A}_{32}\left(\mathbf{A}_{2}^{-1}-\mathbf{A}_{22}\right)^{-1}\mathbf{A}_{21}\right)\\
\times\left(\mathbf{I}-\mathbf{A}_{1}\mathbf{A}_{11}\right)^{-1}\mathbf{c}_1,
\end{multline}
allowing to explicitly express $\mathbf{x}$ in terms of $\mathbf{A}$, $\mathbf{c}_1$, $\mathbf{A}_1$, $\mathbf{A}_2$, and $\mathbf{A}_3$.
In the following, we use this universal framework to derive the channel model based on $Z$-, $Y$-, and $S$-parameters.

\subsection{Impedance Parameters Analysis}
\label{sec:simplified-z}

To simplify \eqref{eq:Hgen-z}, we observe that with a large transmission distance between a transmitter and a receiver, the electrical properties at the transmitter are approximately independent of the electrical properties at the receiver.
The minimum transmission distance at which this phenomenon occurs is a function of the number of antennas at the transmitter and receiver, their radiation pattern, and the antenna spacing, as discussed in \cite{ivr10}.
Thus, assuming that the transmission distance is sufficiently large, we can consider the so-called unilateral approximation and set $\mathbf{Z}_{TI}=\mathbf{0}$, $\mathbf{Z}_{TR}=\mathbf{0}$, and $\mathbf{Z}_{IR}=\mathbf{0}$ \cite{ivr10}.
Note that with this approximation we do not assume the wireless channel to be non-reciprocal.
Conversely, we set to zero the upper block triangular part of $\mathbf{Z}$ as it does not impact the expression of $\mathbf{H}$.
Thus, we can express $\mathbf{v}_T$ and $\mathbf{v}_R$ as
\begin{equation}
\mathbf{v}_T=\left(\mathbf{I}+\mathbf{Z}_{T}\mathbf{Z}_{TT}^{-1}\right)^{-1}\mathbf{v}_{s,T},
\end{equation}
\begin{multline}
\mathbf{v}_R=\mathbf{Z}_{R}\left(\mathbf{Z}_{R}+\mathbf{Z}_{RR}\right)^{-1}
\left(\mathbf{Z}_{RT}-\mathbf{Z}_{RI}\left(\mathbf{Z}_{I}+\mathbf{Z}_{II}\right)^{-1}\mathbf{Z}_{IT}\right)\\
\times\left(\mathbf{Z}_{T}+\mathbf{Z}_{TT}\right)^{-1}\mathbf{v}_{s,T},
\end{multline}
according to Sec.~\ref{sec:simplified-universal}.
Consequently, the channel model based on the $Z$-parameters is given by
\begin{multline}
\mathbf{H}=\mathbf{Z}_{R}\left(\mathbf{Z}_{R}+\mathbf{Z}_{RR}\right)^{-1}\\
\times\left(\mathbf{Z}_{RT}-\mathbf{Z}_{RI}\left(\mathbf{Z}_{I}+\mathbf{Z}_{II}\right)^{-1}\mathbf{Z}_{IT}\right)\mathbf{Z}_{TT}^{-1},\label{eq:Huni-z}
\end{multline}
explicitly clarifying the impact of $\mathbf{Z}$, $\mathbf{Z}_{T}$, $\mathbf{Z}_{I}$, and $\mathbf{Z}_{R}$, in agreement with \cite[Corollary 1]{gra21}.

\subsection{Admittance Parameters Analysis}
\label{sec:simplified-y}

As discussed for the $Z$-parameters, we can also simplify \eqref{eq:Hgen-y} by considering the unilateral approximation.
Setting $\mathbf{Z}_{TI}=\mathbf{0}$, $\mathbf{Z}_{TR}=\mathbf{0}$, and $\mathbf{Z}_{IR}=\mathbf{0}$, and recalling that the entire matrix $\mathbf{Y}$ is related to $\mathbf{Z}$ through $\mathbf{Y}=\mathbf{Z}^{-1}$, we obtain $\mathbf{Y}_{TI}=\mathbf{0}$, $\mathbf{Y}_{TR}=\mathbf{0}$, and $\mathbf{Y}_{IR}=\mathbf{0}$, allowing us to express $\mathbf{i}_T$ and $\mathbf{i}_R$ as
\begin{equation}
\mathbf{i}_T=\left(\mathbf{I}+\mathbf{Y}_{T}\mathbf{Y}_{TT}^{-1}\right)^{-1}\mathbf{i}_{s,T},
\end{equation}
\begin{multline}
\mathbf{i}_R=\mathbf{Y}_{R}\left(\mathbf{Y}_{R}+\mathbf{Y}_{RR}\right)^{-1}\Bigl(\mathbf{Y}_{RT}\Bigr.\\
\left.-\mathbf{Y}_{RI}\left(\mathbf{Y}_{I}+\mathbf{Y}_{II}\right)^{-1}\mathbf{Y}_{IT}\right)
\left(\mathbf{Y}_{T}+\mathbf{Y}_{TT}\right)^{-1}\mathbf{i}_{s,T},
\end{multline}
according to Sec.~\ref{sec:simplified-universal}.
Thus, the channel model based on the $Y$-parameters with the unilateral approximation is given by
\begin{equation}
\mathbf{H}=\left(\mathbf{Y}_{R}+\mathbf{Y}_{RR}\right)^{-1}\left(-\mathbf{Y}_{RT}+\mathbf{Y}_{RI}\left(\mathbf{Y}_{I}+\mathbf{Y}_{II}\right)^{-1}\mathbf{Y}_{IT}\right),\label{eq:Huni-y}
\end{equation}
explicitly defining the effect of $\mathbf{Y}$, $\mathbf{Y}_{T}$, $\mathbf{Y}_{I}$, and $\mathbf{Y}_{R}$.

\subsection{Scattering Parameters Analysis}
\label{sec:simplified-s}

As done for the $Z$- and $Y$-parameters, we simplify \eqref{eq:Hgen-s} by considering the unilateral approximation to better understand the role of the RIS.
Setting $\mathbf{Z}_{TI}=\mathbf{0}$, $\mathbf{Z}_{TR}=\mathbf{0}$, and $\mathbf{Z}_{IR}=\mathbf{0}$, and recalling \eqref{eq:S}, we obtain $\mathbf{S}_{TI}=\mathbf{0}$, $\mathbf{S}_{TR}=\mathbf{0}$, and $\mathbf{S}_{IR}=\mathbf{0}$, which allow us to express $\mathbf{b}_T$ and $\mathbf{b}_R$ as
\begin{equation}
\mathbf{b}_T=\mathbf{S}_{TT}\left(\mathbf{I}-\mathbf{\Gamma}_{T}\mathbf{S}_{TT}\right)^{-1}\mathbf{b}_{s,T},
\end{equation}
\begin{multline}
\mathbf{b}_R=\left(\mathbf{I}-\mathbf{S}_{RR}\mathbf{\Gamma}_{R}\right)^{-1}
\left(\mathbf{S}_{RT}+\mathbf{S}_{RI}\left(\mathbf{I}-\boldsymbol{\Theta}\mathbf{S}_{II}\right)^{-1}\boldsymbol{\Theta}\mathbf{S}_{IT}\right)\\
\times\left(\mathbf{I}-\mathbf{\Gamma}_{T}\mathbf{S}_{TT}\right)^{-1}\mathbf{b}_{s,T},
\end{multline}
according to Sec.~\ref{sec:simplified-universal}.
Thus, the channel model based on the $S$-parameters with the unilateral approximation is given by
\begin{multline}
\mathbf{H}=\left(\mathbf{\Gamma}_{R}+\mathbf{I}\right)\left(\mathbf{I}-\mathbf{S}_{RR}\mathbf{\Gamma}_{R}\right)^{-1}\\
\times\left(\mathbf{S}_{RT}+\mathbf{S}_{RI}\left(\mathbf{I}-\boldsymbol{\Theta}\mathbf{S}_{II}\right)^{-1}\boldsymbol{\Theta}\mathbf{S}_{IT}\right)
\left(\mathbf{I}+\mathbf{S}_{TT}\right)^{-1},\label{eq:Huni-s}
\end{multline}
clearly highlighting the role of $\mathbf{S}$, $\mathbf{\Gamma}_T$, $\boldsymbol{\Theta}$, and $\mathbf{\Gamma}_R$.

Remarkably, the three models \eqref{eq:Huni-z}, \eqref{eq:Huni-y}, and \eqref{eq:Huni-s} effectively show the impact of $\mathbf{Z}_I$, $\mathbf{Y}_I$, and $\boldsymbol{\Theta}$ on \gls{mimo} communication systems, while still capturing the impact of imperfect matching and mutual coupling at the transmitter, RIS, and receiver.

\subsection{Mappings Between Parameters}

Under the unilateral approximation, we can derive simplified mappings that allow us to express the $Y$- and $S$-parameters as a function of the $Z$-parameters.

\subsubsection{From $Z$- to $Y$-Parameters}

To express the matrices $\mathbf{Y}_{RT}$, $\mathbf{Y}_{RI}$, and $\mathbf{Y}_{IT}$ as a function of $\mathbf{Z}$, we consider the relationship $\mathbf{Y}=\mathbf{Z}^{-1}$ where $\mathbf{Z}$ has been simplified through the unilateral approximation, i.e., $\mathbf{Z}_{TI}=\mathbf{0}$, $\mathbf{Z}_{TR}=\mathbf{0}$, and $\mathbf{Z}_{IR}=\mathbf{0}$.
Thus, by computing $\mathbf{Y}$, we have
\begin{equation}
\mathbf{Y}_{RI}=-\mathbf{Z}_{RR}^{-1}\mathbf{Z}_{RI}\mathbf{Z}_{II}^{-1},\;\mathbf{Y}_{IT}=-\mathbf{Z}_{II}^{-1}\mathbf{Z}_{IT}\mathbf{Z}_{TT}^{-1},\label{eq:YRI-YIT-uni}
\end{equation}
\begin{equation}
\mathbf{Y}_{RT}=\mathbf{Z}_{RR}^{-1}\left(-\mathbf{Z}_{RT}+\mathbf{Z}_{RI}\mathbf{Z}_{II}^{-1}\mathbf{Z}_{IT}\right)\mathbf{Z}_{TT}^{-1}.\label{eq:YRT-uni}
\end{equation}

\subsubsection{From $Z$- to $S$-Parameters}

To express the matrices $\mathbf{S}_{RT}$, $\mathbf{S}_{RI}$, and $\mathbf{S}_{IT}$ as a function of $\mathbf{Z}$, we consider the relationship \eqref{eq:S} where $\mathbf{Z}_{TI}=\mathbf{0}$, $\mathbf{Z}_{TR}=\mathbf{0}$, and $\mathbf{Z}_{IR}=\mathbf{0}$.
Thus, it can be proved that
\begin{equation}
\mathbf{S}_{RI}=2Z_0\left(\mathbf{Z}_{RR}+Z_0\mathbf{I}\right)^{-1}\mathbf{Z}_{RI}\left(\mathbf{Z}_{II}+Z_0\mathbf{I}\right)^{-1},\label{eq:SRI-uni}
\end{equation}
\begin{equation}
\mathbf{S}_{IT}=2Z_0\left(\mathbf{Z}_{II}+Z_0\mathbf{I}\right)^{-1}\mathbf{Z}_{IT}\left(\mathbf{Z}_{TT}+Z_0\mathbf{I}\right)^{-1},\label{eq:SIT-uni}
\end{equation}
\begin{multline}
\mathbf{S}_{RT}=2Z_0\left(\mathbf{Z}_{RR}+Z_0\mathbf{I}\right)^{-1}\Bigl(\mathbf{Z}_{RT}\Bigr.\\
\left.-\mathbf{Z}_{RI}\left(\mathbf{Z}_{II}+Z_0\mathbf{I}\right)^{-1}\mathbf{Z}_{IT}\right)\left(\mathbf{Z}_{TT}+Z_0\mathbf{I}\right)^{-1}.\label{eq:SRT-uni}
\end{multline}
Interestingly, these mappings further clarify the equivalence of the three analyses and the meaning of the different terms.
In the following, we show that these mappings play a fundamental role in relating the \gls{em}-consistent models derived in this study with the channel model widely used in related literature.

\begin{table*}[t]
\caption{Channel models based on $Z$-, $Y$-, and $S$-parameters with corresponding assumptions and approximations.}
\resizebox{\textwidth}{!}{
\begin{tabular}{@{}llll@{}}
\toprule
 & $Z$-parameters & $Y$-parameters & $S$-parameters\\
\midrule
Channel model & $\widetilde{\mathbf{Z}}_{RT}\widetilde{\mathbf{Z}}_{TT}^{-1}$ & $\mathbf{Y}_{R}^{-1}\widetilde{\mathbf{Y}}_{RT}\left(\widetilde{\mathbf{Y}}_{TT}-\mathbf{I}\right)^{-1}\mathbf{Y}_{T}$ & $\left(\mathbf{\Gamma}_{R}+\mathbf{I}\right)\widetilde{\mathbf{S}}_{RT}\left(\mathbf{I}+\mathbf{\Gamma}_{T}\widetilde{\mathbf{S}}_{TT}+\widetilde{\mathbf{S}}_{TT}\right)^{-1}$\\
\midrule
A1: Sufficiently large transmission distances$^\star$
& $\mathbf{Z}_{TI}=\mathbf{0}$, $\mathbf{Z}_{TR}=\mathbf{0}$, $\mathbf{Z}_{IR}=\mathbf{0}$
& $\mathbf{Y}_{TI}=\mathbf{0}$, $\mathbf{Y}_{TR}=\mathbf{0}$, $\mathbf{Y}_{IR}=\mathbf{0}$
& $\mathbf{S}_{TI}=\mathbf{0}$, $\mathbf{S}_{TR}=\mathbf{0}$, $\mathbf{S}_{IR}=\mathbf{0}$\\
Channel model with A1
& $\mathbf{Z}_{R}\left(\mathbf{Z}_{R}+\mathbf{Z}_{RR}\right)^{-1}\left(\mathbf{Z}_{RT}-\mathbf{Z}_{RI}\left(\mathbf{Z}_I+\mathbf{Z}_{II}\right)^{-1}\mathbf{Z}_{IT}\right)\mathbf{Z}_{TT}^{-1}$
& $\left(\mathbf{Y}_{R}+\mathbf{Y}_{RR}\right)^{-1}\left(-\mathbf{Y}_{RT}+\mathbf{Y}_{RI}\left(\mathbf{Y}_{I}+\mathbf{Y}_{II}\right)^{-1}\mathbf{Y}_{IT}\right)$
& $\left(\mathbf{\Gamma}_{R}+\mathbf{I}\right)\left(\mathbf{I}-\mathbf{S}_{RR}\mathbf{\Gamma}_{R}\right)^{-1}\left(\mathbf{S}_{RT}+\mathbf{S}_{RI}\left(\mathbf{I}-\boldsymbol{\Theta}\mathbf{S}_{II}\right)^{-1}\boldsymbol{\Theta}\mathbf{S}_{IT}\right)\left(\mathbf{I}+\mathbf{S}_{TT}\right)^{-1}$\\
\midrule
A2: Perfect matching and no mutual coupling at TX and RX
& $\mathbf{Z}_{TT}=Z_0\mathbf{I}$, $\mathbf{Z}_{RR}=Z_0\mathbf{I}$
& $\mathbf{Y}_{TT}=Y_0\mathbf{I}$, $\mathbf{Y}_{RR}=Y_0\mathbf{I}$
& $\mathbf{S}_{TT}=\mathbf{0}$, $\mathbf{S}_{RR}=\mathbf{0}$\\
A3: Impedances at TX and RX equal to $Z_0$
& $\mathbf{Z}_{T}=Z_0\mathbf{I}$, $\mathbf{Z}_{R}=Z_0\mathbf{I}$
& $\mathbf{Y}_{T}=Y_0\mathbf{I}$, $\mathbf{Y}_{R}=Y_0\mathbf{I}$
& $\mathbf{\Gamma}_{T}=\mathbf{0}$, $\mathbf{\Gamma}_{R}=\mathbf{0}$\\
Channel model with A1, A2, and A3
& $\frac{1}{2Z_0}\left(\mathbf{Z}_{RT}-\mathbf{Z}_{RI}\left(\mathbf{Z}_I+\mathbf{Z}_{II}\right)^{-1}\mathbf{Z}_{IT}\right)$
& $\frac{1}{2Y_0}\left(-\mathbf{Y}_{RT}+\mathbf{Y}_{RI}\left(\mathbf{Y}_I+\mathbf{Y}_{II}\right)^{-1}\mathbf{Y}_{IT}\right)$
& $\mathbf{S}_{RT}+\mathbf{S}_{RI}\left(\mathbf{I}-\boldsymbol{\Theta}\mathbf{S}_{II}\right)^{-1}\boldsymbol{\Theta}\mathbf{S}_{IT}$\\
\midrule
A4: Perfect matching and no mutual coupling at RIS
& $\mathbf{Z}_{II}=Z_0\mathbf{I}$
& $\mathbf{Y}_{II}=Y_0\mathbf{I}$
& $\mathbf{S}_{II}=\mathbf{0}$\\
Channel model with A1, A2, A3, and A4
& $\frac{1}{2Z_0}\left(\mathbf{Z}_{RT}-\mathbf{Z}_{RI}\left(\mathbf{Z}_I+Z_0\mathbf{I}\right)^{-1}\mathbf{Z}_{IT}\right)$
& $\frac{1}{2Y_0}\left(-\mathbf{Y}_{RT}+\mathbf{Y}_{RI}\left(\mathbf{Y}_I+Y_0\mathbf{I}\right)^{-1}\mathbf{Y}_{IT}\right)$
& $\mathbf{S}_{RT}+\mathbf{S}_{RI}\boldsymbol{\Theta}\mathbf{S}_{IT}$\\
\bottomrule
\end{tabular}}
\newline\newline
\footnotesize $^\star$ Distances larger than the critical distance provided in \cite{ivr10}.
\label{tab:summary}
\end{table*}

\section{RIS-Aided Communication Model with Perfect Matching and No Mutual Coupling}
\label{sec:model-simplified2}

We have derived three general channel models based on the $Z$-, $Y$-, and $S$-parameters, and we have simplified them by considering the unilateral approximation.
In this section, we further simplify the obtained models to gain engineering insights into the role of RIS in communication systems.

\subsection{Impedance Parameters Analysis}

To further simplify \eqref{eq:Huni-z}, we consider two assumptions in addition to the unilateral approximation.
First, we assume that the antennas at the transmitter and receiver are perfectly matched and have no mutual coupling, yielding $\mathbf{Z}_{TT}=Z_0\mathbf{I}$ and $\mathbf{Z}_{RR}=Z_0\mathbf{I}$.
Second, we assume that the impedances at the transmitter $Z_{T,n_T}$ and the receiver $Z_{R,n_R}$ are all $Z_0$, i.e., $\mathbf{Z}_{T}=Z_0\mathbf{I}$ and $\mathbf{Z}_{R}=Z_0\mathbf{I}$.
With these assumptions, the channel model \eqref{eq:Huni-z} simplifies to
\begin{equation}
\mathbf{H}=\frac{1}{2Z_0}\left(\mathbf{Z}_{RT}-\mathbf{Z}_{RI}\left(\mathbf{Z}_I+\mathbf{Z}_{II}\right)^{-1}\mathbf{Z}_{IT}\right).\label{eq:Hsimp-z-mc}
\end{equation}
In addition, assuming perfect matching and no mutual coupling at the RIS, i.e., $\mathbf{Z}_{II}=Z_0\mathbf{I}$, we obtain
\begin{equation}
\mathbf{H}=\frac{1}{2Z_0}\left(\mathbf{Z}_{RT}-\mathbf{Z}_{RI}\left(\mathbf{Z}_I+Z_0\mathbf{I}\right)^{-1}\mathbf{Z}_{IT}\right),\label{eq:Hsimp-z}
\end{equation}
giving the simplified channel model based on the $Z$-parameters with perfect matching and no mutual coupling.
By further assuming the RIS elements to be canonical minimum scattering antennas \cite{kah65}, the channel without the RIS is equivalent to the channel when the RIS elements are open-circuited, i.e., $\mathbf{Z}_I=\infty\mathbf{I}$, given by $\mathbf{H}=\mathbf{Z}_{RT}/(2Z_0)$ following \eqref{eq:Hsimp-z}.

\subsection{Admittance Parameters Analysis}

To further simplify \eqref{eq:Huni-y}, we consider the two additional assumptions as discussed for the $Z$-parameters.
First, assuming perfect matching and no mutual coupling at the transmitter and receiver, and noting that, with the unilateral approximation, the relationship $\mathbf{Y}=\mathbf{Z}^{-1}$ implies $\mathbf{Y}_{ii}=\mathbf{Z}_{ii}^{-1}$, for $i\in\{T,I,R\}$, we obtain $\mathbf{Y}_{TT}=Y_0\mathbf{I}$ and $\mathbf{Y}_{RR}=Y_0\mathbf{I}$.
Second, considering all the impedances at the transmitter and receiver to be $Z_0$, and recalling that $\mathbf{Y}_{T}=\mathbf{Z}_{T}^{-1}$ and $\mathbf{Y}_{R}=\mathbf{Z}_{R}^{-1}$, we have $\mathbf{Y}_{T}=Y_0\mathbf{I}$ and $\mathbf{Y}_{R}=Y_0\mathbf{I}$.
With these assumptions, the channel model \eqref{eq:Huni-y} simplifies to
\begin{equation}
\mathbf{H}=\frac{1}{2Y_0}\left(-\mathbf{Y}_{RT}+\mathbf{Y}_{RI}\left(\mathbf{Y}_I+\mathbf{Y}_{II}\right)^{-1}\mathbf{Y}_{IT}\right),\label{eq:Hsimp-y-mc}
\end{equation}
In addition, assuming perfect matching and no mutual coupling at the RIS, i.e., $\mathbf{Y}_{II}=Y_0\mathbf{I}$, we obtain
\begin{equation}
\mathbf{H}=\frac{1}{2Y_0}\left(-\mathbf{Y}_{RT}+\mathbf{Y}_{RI}\left(\mathbf{Y}_I+Y_0\mathbf{I}\right)^{-1}\mathbf{Y}_{IT}\right),\label{eq:Hsimp-y}
\end{equation}
giving the simplified channel model based on the $Y$-parameters with perfect matching and no mutual coupling.

\subsection{Scattering Parameters Analysis}

As done for $Z$- and $Y$-parameters, we now further simplify \eqref{eq:Huni-s}.
First, assuming perfect matching and no mutual coupling at the transmitter and receiver, and noting that, with the unilateral approximation, \eqref{eq:S} implies $\mathbf{S}_{ii}=\left(\mathbf{Z}_{ii}+Z_0\mathbf{I}\right)^{-1}\left(\mathbf{Z}_{ii}-Z_0\mathbf{I}\right)$, for $i\in\{T,I,R\}$, we obtain $\mathbf{S}_{TT}=\mathbf{0}$ and $\mathbf{S}_{RR}=\mathbf{0}$.
Second, considering all the impedances at the transmitter and receiver to be $Z_0$, and recalling \eqref{eq:GammaT} and \eqref{eq:GammaR}, we have $\mathbf{\Gamma}_{T}=\mathbf{0}$ and $\mathbf{\Gamma}_{R}=\mathbf{0}$.
With these two assumptions, \eqref{eq:Huni-s} simplifies to
\begin{equation}
\mathbf{H}=\mathbf{S}_{RT}+\mathbf{S}_{RI}\left(\mathbf{I}-\boldsymbol{\Theta}\mathbf{S}_{II}\right)^{-1}\boldsymbol{\Theta}\mathbf{S}_{IT}.\label{eq:Hsimp-s-mc}
\end{equation}
In addition, assuming perfect matching and no mutual coupling at the RIS, i.e., $\mathbf{S}_{II}=\mathbf{0}$, we obtain
\begin{equation}
\mathbf{H}=\mathbf{S}_{RT}+\mathbf{S}_{RI}\boldsymbol{\Theta}\mathbf{S}_{IT},\label{eq:Hsimp-s}
\end{equation}
which is the simplified channel model based on $S$-parameters with perfect matching and no mutual coupling, commonly used in communications in agreement with \cite{she20}.
We summarize the main results of the three analyses based on $Z$-, $Y$-, and $S$-parameters in Tab.~\ref{tab:summary}.

\subsection{Mappings Between Parameters}

Under the assumptions of perfect matching and no mutual coupling, we can derive more simplified mappings that allow us to express the $Y$- and $S$-parameters as a function of the $Z$-parameters.
By using such mappings, it is possible to directly show that the simplified channel models \eqref{eq:Hsimp-z}, \eqref{eq:Hsimp-y}, and \eqref{eq:Hsimp-s} are equivalent.

\subsubsection{From $Z$- to $Y$-Parameters}

Considering the unilateral approximation and perfect matching and no mutual coupling at the transmitter and receiver, the matrices $\mathbf{Y}_{RT}$, $\mathbf{Y}_{RI}$, and $\mathbf{Y}_{IT}$ can be expressed as a function of $\mathbf{Z}$ by setting $\mathbf{Z}_{TT}=Z_0\mathbf{I}$ and $\mathbf{Z}_{RR}=Z_0\mathbf{I}$ in \eqref{eq:YRI-YIT-uni} and \eqref{eq:YRT-uni}, yielding
\begin{equation}
\mathbf{Y}_{RI}=-\frac{\mathbf{Z}_{RI}\mathbf{Z}_{II}^{-1}}{Z_0},\;\mathbf{Y}_{IT}=-\frac{\mathbf{Z}_{II}^{-1}\mathbf{Z}_{IT}}{Z_0},\label{eq:YRI-YIT-mc}
\end{equation}
\begin{equation}
\mathbf{Y}_{RT}=\frac{1}{Z_0^2}\left(-\mathbf{Z}_{RT}+\mathbf{Z}_{RI}\mathbf{Z}_{II}^{-1}\mathbf{Z}_{IT}\right).\label{eq:YRT-mc}
\end{equation}
In addition, with perfect matching and no mutual coupling at the RIS, i.e., $\mathbf{Z}_{II}=Z_0\mathbf{I}$, \eqref{eq:YRI-YIT-mc} and \eqref{eq:YRT-mc} boil down to
\begin{equation}
\mathbf{Y}_{RI}=-\frac{\mathbf{Z}_{RI}}{Z_0^2},\;\mathbf{Y}_{IT}=-\frac{\mathbf{Z}_{IT}}{Z_0^2},\label{eq:YRI-YIT}
\end{equation}
\begin{equation}
\mathbf{Y}_{RT}=\frac{1}{Z_0^2}\left(-\mathbf{Z}_{RT}+\frac{\mathbf{Z}_{RI}\mathbf{Z}_{IT}}{Z_0}\right).\label{eq:YRT}
\end{equation}
Interestingly, by substituting $\mathbf{Y}_{I}=\mathbf{Z}_{I}^{-1}$, \eqref{eq:YRI-YIT}, and \eqref{eq:YRT} into \eqref{eq:Hsimp-y}, we directly obtain \eqref{eq:Hsimp-z}, confirming that the two analyses based on $Z$- and $Y$-parameters lead to the same result.

\subsubsection{From $Z$- to $S$-Parameters}

To express the matrices $\mathbf{S}_{RT}$, $\mathbf{S}_{RI}$, and $\mathbf{S}_{IT}$ as a function of $\mathbf{Z}$ with the unilateral approximation and perfect matching and no mutual coupling at the transmitter and receiver, we can set $\mathbf{Z}_{TT}=Z_0\mathbf{I}$ and $\mathbf{Z}_{RR}=Z_0\mathbf{I}$ in \eqref{eq:SRI-uni}, \eqref{eq:SIT-uni}, and \eqref{eq:SRT-uni}, yielding
\begin{equation}
\mathbf{S}_{RI}=\mathbf{Z}_{RI}\left(\mathbf{Z}_{II}+Z_0\mathbf{I}\right)^{-1},\;\mathbf{S}_{IT}=\left(\mathbf{Z}_{II}+Z_0\mathbf{I}\right)^{-1}\mathbf{Z}_{IT},\label{eq:SRI-SIT-mc}
\end{equation}
\begin{equation}
\mathbf{S}_{RT}=\frac{1}{2Z_0}\left(\mathbf{Z}_{RT}-\mathbf{Z}_{RI}\left(\mathbf{Z}_{II}+Z_0\mathbf{I}\right)^{-1}\mathbf{Z}_{IT}\right).\label{eq:SRT-mc}
\end{equation}
In addition, with perfect matching and no mutual coupling at the RIS, i.e., $\mathbf{Z}_{II}=Z_0\mathbf{I}$, \eqref{eq:SRI-SIT-mc} and \eqref{eq:SRT-mc} boil down to
\begin{equation}
\mathbf{S}_{RI}=\frac{\mathbf{Z}_{RI}}{2Z_0},\;\mathbf{S}_{IT}=\frac{\mathbf{Z}_{IT}}{2Z_0},\label{eq:SRI-SIT}
\end{equation}
\begin{equation}
\mathbf{S}_{RT}=\frac{1}{2Z_0}\left(\mathbf{Z}_{RT}-\frac{\mathbf{Z}_{RI}\mathbf{Z}_{IT}}{2Z_0}\right).\label{eq:SRT}
\end{equation}
Interestingly, by substituting \eqref{eq:Theta}, \eqref{eq:SRI-SIT}, and \eqref{eq:SRT} into \eqref{eq:Hsimp-s}, we directly obtain \eqref{eq:Hsimp-z}, confirming that the two analyses based on the $Z$- and $S$-parameters lead to the same conclusion.

\subsection{Relationship with the Widely Used Model}

We have derived three equivalent simplified channel models under the assumptions of perfect matching and no mutual coupling.
In this section, we relate these models with the RIS-aided channel model widely used in related literature.

In RIS literature \cite{wu19b}-\cite{li22-3}, the notation
\begin{equation}
\mathbf{H}_{RT}=\mathbf{S}_{RT},\;\mathbf{H}_{RI}=\mathbf{S}_{RI},\;\mathbf{H}_{IT}=\mathbf{S}_{IT},\label{eq:Hij}
\end{equation}
is commonly used to denote the channel matrices from the transmitter to the receiver, from the RIS to the receiver, and from the transmitter to the RIS, respectively.
By substituting \eqref{eq:Hij} into the simplified channel based on the $S$-parameters \eqref{eq:Hsimp-s}, we obtain
\begin{equation}
\mathbf{H}=\mathbf{H}_{RT}+\mathbf{H}_{RI}\boldsymbol{\Theta}\mathbf{H}_{IT},\label{eq:H}
\end{equation}
which is the RIS-aided communication model widely used in related literature.
Thus, we can conclude that the widely used model in \eqref{eq:H} is derived by considering the unilateral approximation and under the assumptions of perfect matching and no mutual coupling, in agreement with \cite{she20}.

An interesting aspect of the channel model \eqref{eq:H} is the dependence of $\mathbf{H}_{RT}$ on $\mathbf{H}_{RI}$ and $\mathbf{H}_{IT}$ as a result of \eqref{eq:SRI-SIT}, \eqref{eq:SRT}, and \eqref{eq:Hij}.
However, this aspect is commonly neglected in related literature, where it is implicitly considered
\begin{equation}
\mathbf{S}_{RT}\approx\frac{\mathbf{Z}_{RT}}{2Z_0},\label{eq:SRT-approx}
\end{equation}
as an approximation of \eqref{eq:SRT}.
Remarkably, \eqref{eq:SRT-approx} can be obtained by considering the relationship
\begin{equation}
\mathbf{Z}=2Z_0\left(\mathbf{I}-\mathbf{S}\right)^{-1}-Z_0\mathbf{I},\label{eq:Z}
\end{equation}
which is derived from \eqref{eq:S}, by applying the first-order Neumann series approximation $\left(\mathbf{I}-\mathbf{S}\right)^{-1}\approx\left(\mathbf{I}+\mathbf{S}\right)$.
With this approximation, from \eqref{eq:Z} we obtain $\mathbf{Z}\approx Z_0\left(\mathbf{I}+2\mathbf{S}\right)$, directly giving \eqref{eq:SRT-approx}.
Note that the first-order Neumann series approximation holds as long as $\Vert\mathbf{S}\Vert_F$ is small, which is typically valid in wireless communication scenarios given the high losses suffered from the propagating signal.


According to the approximation in \eqref{eq:SRT-approx}, when the direct channel between the transmitter and receiver is completely obstructed, i.e., $\mathbf{Z}_{RT}=\mathbf{0}$, we have $\mathbf{H}_{RT}=\mathbf{0}$, as widely accepted in related literature.
However, according to \eqref{eq:SRT}, a completely obstructed direct channel, i.e., $\mathbf{Z}_{RT}=\mathbf{0}$, gives $\mathbf{H}_{RT}=-\mathbf{H}_{RI}\mathbf{H}_{IT}$, which is in general non zero, as correctly noticed in \cite{abr23,nos23}.
The physical meaning of having $\mathbf{H}_{RT}\neq\mathbf{0}$ in the case of a completely obstructed direct channel, i.e., $\mathbf{Z}_{RT}=\mathbf{0}$, lies in the structural scattering of the RIS.

Since the approximation in \eqref{eq:SRT-approx} causes an inconsistency, especially when $\mathbf{Z}_{RT}=\mathbf{0}$, it is worth quantifying the difference between \eqref{eq:H} and the channel model resulting from \eqref{eq:SRT-approx}, which is widely used in the literature.
To this end, we assume a completely obstructed direct channel, i.e., $\mathbf{Z}_{RT}=\mathbf{0}$, and consider a \gls{siso} system, i.e., $N_R=1$ and $N_T=1$, where \eqref{eq:H} boils down to
\begin{equation}
h=-\mathbf{h}_{RI}\mathbf{h}_{IT}+\mathbf{h}_{RI}\boldsymbol{\Theta}\mathbf{h}_{IT},\label{eq:h}
\end{equation}
and, considering the approximation in \eqref{eq:SRT-approx}, to
\begin{equation}
h\approx\mathbf{h}_{RI}\boldsymbol{\Theta}\mathbf{h}_{IT}\triangleq h^\prime.\label{eq:hprime}
\end{equation}
Interestingly, in \eqref{eq:h}, the structural scattering term $-\mathbf{h}_{RI}\mathbf{h}_{IT}$ can constructively or destructively interfere with $\mathbf{h}_{RI}\boldsymbol{\Theta}\mathbf{h}_{IT}$ depending on $\boldsymbol{\Theta}$, e.g., when $\boldsymbol{\Theta}=-\mathbf{I}$ or $\boldsymbol{\Theta}=\mathbf{I}$, respectively.
This is because the structural scattering term mainly results in a reflection towards the specular direction, while the direction of reflection of the second term mainly depends on the RIS scattering matrix.
In the following, we derive the scaling laws of the received signal power achieved under the channel models in \eqref{eq:h} and \eqref{eq:hprime} with a conventional RIS architecture.

As for the model in \eqref{eq:h}, the received signal power is given by $P_R=\vert-\mathbf{h}_{RI}\mathbf{h}_{IT}+\mathbf{h}_{RI}\boldsymbol{\Theta}\mathbf{h}_{IT}\vert^2$, where a unit transmit power is assumed with no loss of generality.
Thus, the maximum received signal power is given by
\begin{equation}
P_R=\left(\left\vert\mathbf{h}_{RI}\mathbf{h}_{IT}\right\vert+\sum_{n_I=1}^{N_I}\left\vert\left[\mathbf{h}_{RI}\right]_{n_I}\left[\mathbf{h}_{IT}\right]_{n_I}\right\vert\right)^2,\label{eq:PR}
\end{equation}
achievable by co-phasing the terms $-\mathbf{h}_{RI}\mathbf{h}_{IT}$ and $\mathbf{h}_{RI}\boldsymbol{\Theta}\mathbf{h}_{IT}$ through an appropriate phase shift introduced by $\boldsymbol{\Theta}$.

As for the widely used model $h^\prime$, the received signal power is equal to $P_R^\prime=\vert\mathbf{h}_{RI}\boldsymbol{\Theta}\mathbf{h}_{IT}\vert^2$.
Thus, it is possible to optimize the RIS to achieve a received signal power of
\begin{equation}
P_R^\prime=\left(\sum_{n_I=1}^{N_I}\left\vert\left[\mathbf{h}_{RI}\right]_{n_I}\left[\mathbf{h}_{IT}\right]_{n_I}\right\vert\right)^2.\label{eq:PRprime}
\end{equation}
We define the relative difference between the average received signal power $\text{E}[P_R]$ and $\text{E}[P_R^\prime]$ as
\begin{equation}
\delta=\frac{\text{E}\left[P_R\right]-\text{E}\left[P_R^\prime\right]}{\text{E}\left[P_R\right]}.\label{eq:delta}
\end{equation}

\begin{figure}[t]
\centering
\includegraphics[width=0.4\textwidth]{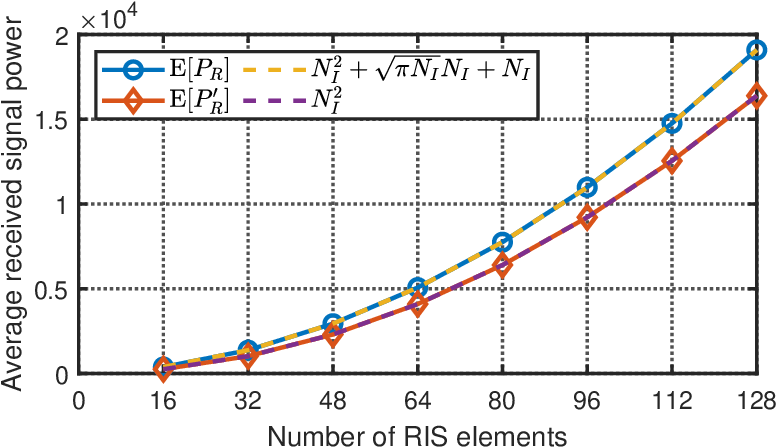}
\caption{Average received signal power obtained with the model $h$ and the widely used model $h^\prime$.}
\label{fig:perf}
\end{figure}
\begin{figure}[t]
\centering
\includegraphics[width=0.4\textwidth]{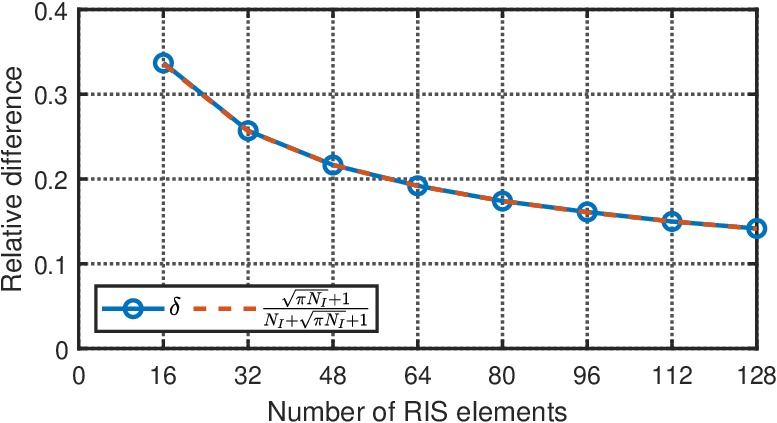}
\caption{Relative difference between the average received signal power obtained with the model $h$ and the widely used model $h^\prime$.}
\label{fig:diff}
\end{figure}

To gain numerical insights into \eqref{eq:delta}, we consider \gls{los} channels given by $\mathbf{h}_{RI}=[e^{\phi_1},\ldots,e^{\phi_{N_I}}]$ and $\mathbf{h}_{IT}=[e^{\psi_1},\ldots,e^{\psi_{N_I}}]^T$.
With these channels, in the case of the model $h$, we have
\begin{equation}
\text{E}\left[P_R\right]=\text{E}\left[\left\vert\mathbf{h}_{RI}\mathbf{h}_{IT}\right\vert^2\right]+N_I^2+2N_I\text{E}\left[\left\vert\mathbf{h}_{RI}\mathbf{h}_{IT}\right\vert\right].
\end{equation}
Approximating the term $\mathbf{h}_{RI}\mathbf{h}_{IT}$ as a sum of $N_I$ \gls{iid} random variables with mean $0$ and variance $1$, it holds $\mathbf{h}_{RI}\mathbf{h}_{IT}\sim\mathcal{CN}(0,N_I)$ by the Central Limit Theorem.
Thus, by using $\text{E}[\vert\mathbf{h}_{RI}\mathbf{h}_{IT}\vert]=\sqrt{\frac{\pi}{4}N_I}$ and $\text{E}[\vert\mathbf{h}_{RI}\mathbf{h}_{IT}\vert^2]=N_I$, we have
\begin{equation}
\text{E}\left[P_R\right]=N_I^2+\sqrt{\pi N_I}N_I+N_I.\label{eq:EPR-los}
\end{equation}
Besides, in the case of the widely used model $h^\prime$, we have
\begin{equation}
\text{E}\left[P_R^\prime\right]=P_R^\prime=N_I^2.\label{eq:PRprime-los}
\end{equation}
Considering \eqref{eq:EPR-los} and \eqref{eq:PRprime-los}, the relative difference between $\text{E}[P_R]$ and $\text{E}[P_R^\prime]$ under \gls{los} channels is equal to
\begin{equation}
\delta=\frac{\sqrt{\pi N_I}+1}{N_I+\sqrt{\pi N_I}+1},
\end{equation}
giving $\delta\rightarrow0$ for $N_I\rightarrow\infty$.

In Fig.~\ref{fig:perf}, we report the average received signal power achieved with $h$ and the approximated model $h^\prime$.
We observe that the received signal power for the channel $h$ is higher because of the additional term $-\mathbf{h}_{RI}\mathbf{h}_{IT}$.
In addition, in Fig.~\ref{fig:diff}, we report the relative difference between the average received signal power achieved with $h$ and with the approximated model $h^\prime$.
We observe that, despite the difference vanishes when $N_I\rightarrow\infty$, it is non-negligible for a practical number of RIS elements since it vanishes slowly with $N_I$.

\section{RIS Architectures and Optimization}
\label{sec:architecture-optimization}

We have shown that the $Z$-, $Y$-, and $S$-parameters can be equivalently used to characterize an RIS-aided communication model.
Similarly, they can be equivalently used to characterize the $N_I$-port reconfigurable impedance network implementing the RIS.
In this section, we discuss how the matrices $\mathbf{Z}_I$, $\mathbf{Y}_I$, and $\boldsymbol{\Theta}$ can be used to characterize different RIS architectures, including conventional RIS and BD-RIS.
In addition, we discuss the advantages of the $Z$-, $Y$-, and $S$-parameters in optimizing an RIS.

\subsection{RIS Architectures}

In conventional RIS architectures, also known as single-connected RIS, each RIS port is solely connected to the ground through a tunable impedance \cite{she20}.
Hence, the resulting impedance, admittance, and scattering matrices are diagonal, as given in Tab.~\ref{tab:architectures}, where $Z_{n_I}$ is the tunable impedance connecting the $n_I$th RIS port to the ground, $Y_{n_I}=Z_{n_I}^{-1}$, and $\Theta_{n_I}$ is the reflection coefficient corresponding to $Z_{n_I}$, given by $\Theta_{n_I}=(Z_{n_I}-Z_0)/(Z_{n_I}+Z_0)$, according to \eqref{eq:Theta}, for $n_I=1,\ldots,N_I$.
Furthermore, in the case of a lossless single-connected RIS, $Z_{n_I}$ and $Y_{n_I}$ are purely imaginary, and $\vert\Theta_{n_I}\vert=1$, for $n_I=1,\ldots,N_I$, as shown in Tab.~\ref{tab:architectures}.
This conventional circuit topology is the simplest RIS circuit topology, offering limited flexibility \cite{she20}.

\begin{table*}[t]
\centering
\caption{Constraints of lossless and reciprocal RIS architectures and applications based on $Z$-, $Y$-, and $S$-parameters.}
\resizebox{\textwidth}{!}{
\begin{tabular}{@{}llll@{}}
\toprule
&$Z$-parameters & $Y$-parameters & $S$-parameters\\
\midrule
Single-connected
&\begin{tabular}{@{}l@{}}$\mathbf{Z}_I=\mathrm{diag}\left(Z_{1},\ldots,Z_{N_I}\right)$,\\
$Z_{n_I}=jX_{n_I}$, $X_{n_I}\in\mathbb{R}$, $\forall n_I$\end{tabular}
&\begin{tabular}{@{}l@{}}$\mathbf{Y}_I=\mathrm{diag}\left(Y_{1},\ldots,Y_{N_I}\right)$,\\
$Y_{n_I}=jB_{n_I}$, $B_{n_I}\in\mathbb{R}$, $\forall n_I$\end{tabular}
&\begin{tabular}{@{}l@{}}$\boldsymbol{\Theta}=\mathrm{diag}\left(\Theta_{1},\ldots,\Theta_{N_I}\right)$,\\
$\Theta_{n_I}=e^{j\theta_{n_I}}$, $\theta_{n_I}\in[0,2\pi)$, $\forall n_I$\end{tabular}\\
\midrule
Fully-connected
&$\mathbf{Z}_{I}=j\mathbf{X}_{I}$, $\mathbf{X}_{I}=\mathbf{X}_{I}^T$, $\mathbf{X}_{I}\in\mathbb{R}^{N_I\times N_I}$
&$\mathbf{Y}_{I}=j\mathbf{B}_{I}$, $\mathbf{B}_{I}=\mathbf{B}_{I}^T$, $\mathbf{B}_{I}\in\mathbb{R}^{N_I\times N_I}$
&$\boldsymbol{\Theta}^H\boldsymbol{\Theta}=\mathbf{I}$, $\boldsymbol{\Theta}=\boldsymbol{\Theta}^T$\\
\midrule
Group-connected
&\begin{tabular}{@{}l@{}}$\mathbf{Z}_{I}=\mathrm{diag}\left(\mathbf{Z}_{I,1},\ldots,\mathbf{Z}_{I,G}\right)$,\\
$\mathbf{Z}_{I,g}=j\mathbf{X}_{I,g}$, $\mathbf{X}_{I,g}=\mathbf{X}_{I,g}^T$, $\mathbf{X}_{I,g}\in\mathbb{R}^{N_G\times N_G}$, $\forall g$\end{tabular}
&\begin{tabular}{@{}l@{}}$\mathbf{Y}_{I}=\mathrm{diag}\left(\mathbf{Y}_{I,1},\ldots,\mathbf{Y}_{I,G}\right)$,\\
$\mathbf{Y}_{I,g}=j\mathbf{B}_{I,g}$, $\mathbf{B}_{I,g}=\mathbf{B}_{I,g}^T$, $\mathbf{B}_{I,g}\in\mathbb{R}^{N_G\times N_G}$, $\forall g$\end{tabular}
&\begin{tabular}{@{}l@{}}$\boldsymbol{\Theta}=\mathrm{diag}\left(\boldsymbol{\Theta}_{1},\ldots,\boldsymbol{\Theta}_{G}\right)$,\\
$\boldsymbol{\Theta}_{g}^H\boldsymbol{\Theta}_{g}=\mathbf{I}$, $\boldsymbol{\Theta}_g=\boldsymbol{\Theta}_g^T$, $\forall g$\end{tabular}\\
\midrule
Tree-connected$^\star$
&$-$
&\begin{tabular}{@{}l@{}}$\mathbf{Y}_{I}=j\mathbf{B}_{I}$, $\mathbf{B}_{I}=\mathbf{B}_{I}^T$, $\mathbf{B}_{I}\in\mathbb{R}^{N_I\times N_I}$,\\
$\left[\mathbf{B}_{I}\right]_{i,j}=0$ if $\left\vert i-j\right\vert>1$\end{tabular}
&$-$\\
\midrule
Forest-connected$^\star$
&$-$
&\begin{tabular}{@{}l@{}}$\mathbf{Y}_{I}=\mathrm{diag}\left(\mathbf{Y}_{I,1},\ldots,\mathbf{Y}_{I,G}\right)$,\\
$\mathbf{Y}_{I,g}=j\mathbf{B}_{I,g}$, $\mathbf{B}_{I,g}=\mathbf{B}_{I,g}^T$, $\mathbf{B}_{I,g}\in\mathbb{R}^{N_G\times N_G}$,\\
$\left[\mathbf{B}_{I,g}\right]_{i,j}=0$ if $\left\vert i-j\right\vert>1$, $\forall g$\end{tabular} 
&$-$\\
\midrule
Application
&Optimization w/ mutual coupling
&Optimization of BD-RIS with sparse $\mathbf{Y}_{I}$
&Optimization w/o mutual coupling\\
\bottomrule
\end{tabular}}
\newline\newline
\footnotesize $^\star$ For these RIS families, the tridiagonal RIS architecture is considered \cite{ner23-1}, and the constraints cannot be expressed in terms of the $Z$- and $S$-parameters.
\label{tab:architectures}
\end{table*}

To overcome the limited flexibility of conventional RIS architectures, the ports in BD-RIS can also be connected to each other through additional tunable impedance (or admittance) components.
By denoting the tunable admittance connecting the $n_I$th port to the $m_I$th port as $Y_{n_I,m_I}$, the $(n_I,m_I)$th entry of the admittance matrix is given by 
\begin{equation}
\left[\mathbf{Y}_I\right]_{n_I,m_I}=
\begin{cases}
-Y_{n_I,m_I} & n_I\neq m_I\\
Y_{n_I}+\sum_{k\neq n_I}Y_{n_I,k} & n_I=m_I
\end{cases},\label{eq:Yij}
\end{equation}
as discussed in \cite{she20}.
Thus, the admittance matrix of BD-RIS is not limited to being diagonal.
The same holds for the impedance and scattering matrices of BD-RIS, as it can be easily noticed by applying $\mathbf{Z}_I=\mathbf{Y}_I^{-1}$ and \eqref{eq:Theta}.
Depending on the topology of the interconnections between the RIS ports, multiple RIS architectures have been proposed.
In Tab.~\ref{tab:architectures}, we report the constraint of fully-/group-connected RIS \cite{she20}, and tree-/forest-connected RIS with the tridiagonal architecture \cite{ner23-1} based on the three parameters.
In group- and forest-connected RIS, the elements are grouped into $G$ groups, each with size $N_G=N/G$, such that $\mathbf{Z}_{I,g}\in\mathbb{C}^{N_G\times N_G}$, $\mathbf{Y}_{I,g}\in\mathbb{C}^{N_G\times N_G}$, and $\boldsymbol{\Theta}_{g}\in\mathbb{C}^{N_G\times N_G}$ are the impedance, admittance, and scattering matrix of the $g$th group, for $g=1,\ldots,G$, respectively \cite{she20,ner23-1}.
The $N_I$-port network of the RIS is often assumed to be reciprocal (not including non-reciprocal media) and lossless (to maximize the scattered power).
If the RIS is reciprocal, the impedance, admittance, and scattering matrices are symmetric.
Furthermore, if the RIS is lossless, the impedance and admittance matrices are purely imaginary while the scattering matrix is unitary, as shown in Tab.~\ref{tab:architectures}.

\subsection{RIS Optimization Based on the Impedance Parameters}
\label{sec:opt-z}

Since the $Z$-, $Y$-, and $S$-parameters can be used interchangeably to characterize an RIS architecture, we now highlight the advantages of using the different parameters to optimize an RIS.
To this end, we consider three case studies derived from the received signal power maximization problem in an RIS-aided \gls{mimo} system.
In the considered communication system, the transmitted signal is expressed as $\mathbf{v}_T=\mathbf{w}s$, where $\mathbf{w}\in\mathbb{C}^{N_{T}\times1}$ is the normalized precoder subject to $\Vert\mathbf{w}\Vert=1$, and $s\in\mathbb{C}$ is the transmitted symbol with average power $P_{T}=\mathrm{E}[\vert s\vert^{2}]$.
Besides, the signal used for detection is given by $z=\mathbf{g}\mathbf{v}_R\in\mathbb{C}$, where $\mathbf{g}\in\mathbb{C}^{1\times N_{R}}$ is the normalized combiner subject to the constraint $\Vert\mathbf{g}\Vert=1$, and $\mathbf{v}_R=\mathbf{H}\mathbf{v}_T$ is the received signal.
Thus, the received signal power is given by $P_R=P_T\vert\mathbf{g}\mathbf{H}\mathbf{w}\vert^{2}$.

The $Z$-parameters are proved to be the effective representation when optimizing an RIS in the presence of mutual coupling \cite{qia21}-\cite{akr23}.
Specifically, an RIS characterized through the $Z$-parameters has been optimized by applying the Neumann series approximation \cite{qia21}-\cite{mur23}, by iteratively optimizing the tunable loads in closed-form \cite{has23}, and through the gradient ascent algorithm \cite{akr23}.
Furthermore, it has been shown that the $Z$-parameters facilitate the optimization of BD-RIS in the presence of mutual coupling at the RIS \cite{li23-3}.

As a case study, we consider the received signal power maximization problem in an RIS-aided system in the presence of mutual coupling, with the RIS being group-connected, including single- and fully-connected as two special cases \cite{she20}.
Considering a lossless and reciprocal RIS, i.e., with $\mathbf{Z}_{I}=j\mathbf{X}_{I}$ and $\mathbf{X}_{I}=\mathbf{X}_{I}^T$, where $\mathbf{X}_{I}$ is the RIS reactance matrix, and $\mathbf{H}$ is given in \eqref{eq:Hsimp-z-mc}, our problem writes as
\begin{align}
\underset{\mathbf{w},\mathbf{g},\mathbf{X}_I}{\mathsf{\mathrm{max}}}\;\;
&\frac{P_{T}}{4Z_0^2}\left\vert\mathbf{g}\left(\mathbf{Z}_{RT}-\mathbf{Z}_{RI}\left(j\mathbf{X}_{I}+\mathbf{Z}_{II}\right)^{-1}\mathbf{Z}_{IT}\right)\mathbf{w}\right\vert^{2}\label{eq:PR-z}\\
\mathsf{\mathrm{s.t.}}\;\;\;
&\mathbf{X}_{I}=\mathrm{diag}\left(\mathbf{X}_{I,1},\ldots,\mathbf{X}_{I,G}\right),\label{eq:PR-z-c1}\\
&\mathbf{X}_{I,g}=\mathbf{X}_{I,g}^{T},\;\forall g,\label{eq:PR-z-c2}\\
&\left\Vert\mathbf{w}\right\Vert=1,\;\left\Vert\mathbf{g}\right\Vert=1,\label{eq:PR-z-c3}
\end{align}
which is solved by jointly optimizing $\mathbf{w}$, $\mathbf{g}$, and $\mathbf{X}_I$.
To solve \eqref{eq:PR-z}-\eqref{eq:PR-z-c3}, we initialize $\mathbf{X}_I$ to a feasible value and alternate between the following two steps.
First, with $\mathbf{X}_I$ fixed, $\mathbf{w}$ and $\mathbf{g}$ are updated as the dominant right and left singular vectors of $\mathbf{Z}_{RT}-\mathbf{Z}_{RI}(j\mathbf{X}_I+\mathbf{Z}_{II})^{-1}\mathbf{Z}_{IT}$, respectively, which is globally optimal.
Second, with $\mathbf{w}$ and $\mathbf{g}$ fixed, $\mathbf{X}_I$ is updated by solving
\begin{equation}
\underset{\mathbf{X}_I}{\mathsf{\mathrm{max}}}\;
\left\vert z_{RT}^{\text{eff}}-\mathbf{z}_{RI}^{\text{eff}}\left(j\mathbf{X}_I+\mathbf{Z}_{II}\right)^{-1}\mathbf{z}_{IT}^{\text{eff}}\right\vert^{2}\;
\mathsf{\mathrm{s.t.}}\;
\eqref{eq:PR-z-c1},\;\eqref{eq:PR-z-c2},\label{eq:PR-z2}
\end{equation}
where $z_{RT}^{\text{eff}}=\mathbf{g}\mathbf{Z}_{RT}\mathbf{w}$, $\mathbf{z}_{RI}^{\text{eff}}=\mathbf{g}\mathbf{Z}_{RI}$, and $\mathbf{z}_{IT}^{\text{eff}}=\mathbf{Z}_{IT}\mathbf{w}$, as proposed in \cite{li23-3}.
These two steps are alternatively repeated until convergence of the objective function in \eqref{eq:PR-z}.

\subsection{RIS Optimization Based on the Admittance Parameters}
\label{sec:opt-y}

The optimization of an RIS based on the $Y$-parameters is convenient when considering BD-RIS architectures whose admittance matrix $\mathbf{Y}_I$ is sparse, i.e., there is only a limited number of tunable admittance components interconnecting the RIS elements, like in tree- and forest-connected RISs \cite{ner23-1}.
In this case, the $Y$-parameters are an effective representation since the entries of $\mathbf{Y}_I$ are directly linked to the tunable admittance components in the BD-RIS circuit topology, as given by \eqref{eq:Yij}.

As a case study, we consider the received signal power maximization problem in an RIS-aided system, with the RIS being forest-connected, including single- and tree-connected as two special cases \cite{ner23-1}.
Considering lossless and reciprocal RIS, i.e., with $\mathbf{Y}_{I}=j\mathbf{B}_{I}$ and $\mathbf{B}_{I}=\mathbf{B}_{I}^T$, where $\mathbf{B}_{I}$ is the RIS susceptance matrix, and $\mathbf{H}$ is given in \eqref{eq:Hsimp-y}, assuming no mutual coupling, our problem writes as
\begin{align}
\underset{\mathbf{w},\mathbf{g},\mathbf{B}_I}{\mathsf{\mathrm{max}}}\;\;
&\frac{P_{T}}{4Y_0^2}\left|\mathbf{g}\left(-\mathbf{Y}_{RT}+\mathbf{Y}_{RI}\left(j\mathbf{B}_I+Y_0\mathbf{I}\right)^{-1}\mathbf{Y}_{IT}\right)\mathbf{w}\right|^{2}\label{eq:PR-y}\\
\mathsf{\mathrm{s.t.}}\;\;\;
&\mathbf{B}_I=\mathrm{diag}\left(\mathbf{B}_{I,1},\ldots,\mathbf{B}_{I,G}\right),\label{eq:PR-y-c1}\\
&\mathbf{B}_{I,g}=\mathbf{B}_{I,g}^{T},\:\left[\mathbf{B}_{I,g}\right]_{i,j}=0\textrm{ if }\left\vert i-j\right\vert>1,\forall g,\label{eq:PR-y-c2}\\
&\left\|\mathbf{w}\right\|=1,\;\left\|\mathbf{g}\right\|=1,\label{eq:PR-y-c3}
\end{align}
where \eqref{eq:PR-y-c2} indicates that the $g$th block of the susceptance matrix $\mathbf{B}_{g}$ has a tridiagonal architecture, as introduced in \cite{ner23-1}.
To jointly optimize $\mathbf{w}$, $\mathbf{g}$, and $\mathbf{B}_I$, we initialize $\mathbf{B}_I$ to a feasible value and alternate between the following two steps.
First, with $\mathbf{B}_I$ fixed, $\mathbf{w}$ and $\mathbf{g}$ are updated as the dominant right and left singular vectors of $-\mathbf{Y}_{RT}+\mathbf{Y}_{RI}(j\mathbf{B}_I+Y_0\mathbf{I})^{-1}\mathbf{Y}_{IT}$, respectively, which is a global optimal solution.
Second, with $\mathbf{w}$ and $\mathbf{g}$ fixed, $\mathbf{B}_I$ is updated by solving
\begin{equation}
\underset{\mathbf{B}_I}{\mathsf{\mathrm{max}}}\;
\left\vert-y_{RT}^{\text{eff}}+\mathbf{y}_{RI}^{\text{eff}}\left(j\mathbf{B}_I+Y_0\mathbf{I}\right)^{-1}\mathbf{y}_{IT}^{\text{eff}}\right\vert^{2}\;
\mathsf{\mathrm{s.t.}}\;
\eqref{eq:PR-y-c1},\;\eqref{eq:PR-y-c2},\label{eq:PR-y2}
\end{equation}
where $y_{RT}^{\text{eff}}=\mathbf{g}\mathbf{Y}_{RT}\mathbf{w}$, $\mathbf{y}_{RI}^{\text{eff}}=\mathbf{g}\mathbf{Y}_{RI}$, and $\mathbf{y}_{IT}^{\text{eff}}=\mathbf{Y}_{IT}\mathbf{w}$.
Remarkably, \eqref{eq:PR-y2} has been solved in \cite{ner23-1}, where a globally optimal solution for each block $\mathbf{B}_{I,g}$ has been proposed.
These two steps are alternatively repeated until convergence of the objective function in \eqref{eq:PR-y}.

\subsection{RIS Optimization Based on the Scattering Parameters}
\label{sec:opt-s}

The $S$-parameters model has been used to optimize conventional RIS in the vast majority of related works \cite{wu19b}-\cite{che22}, where perfect matching and no mutual coupling are implicitly assumed.
The $S$-parameters model owns its popularity to the widely used channel model in \eqref{eq:Hsimp-s}, in which $\boldsymbol{\Theta}=\mathrm{diag}\left(e^{j\theta_{1}},\ldots,e^{j\theta_{N_I}}\right)$ for a lossless single-connected RIS architecture, and due to its role played in channel measurement \cite{gui14}.
For single-connected RIS, the $S$-parameters allow to directly optimize the phase shifts $\theta_{n_I}\in[0,2\pi)$ through several optimization techniques, including \gls{sdr} \cite{wu19b}, iterative closed-form solutions \cite{guo20}, \gls{mm} and manifold optimization \cite{pan20}, \gls{admm} \cite{liu22}, and \gls{bb} methods \cite{di20}.
For BD-RIS, the $S$-parameters enable the use of manifold optimization by exploiting the unitary constraint on $\boldsymbol{\Theta}$ \cite{li22-1}-\cite{li23-2}.
In addition, the $S$-parameters allow for efficient optimization of BD-RIS by using tailored decompositions of $\boldsymbol{\Theta}$ \cite{ner22}, through symmetric unitary projection \cite{fan23}, and by solving the orthogonal Procrustes problem \cite{wan23}.

\begin{figure*}[t]
\centering
\subfigure[$Z$-parameters, w/ and w/o mutual coupling.]{
\includegraphics[width=0.4\textwidth]{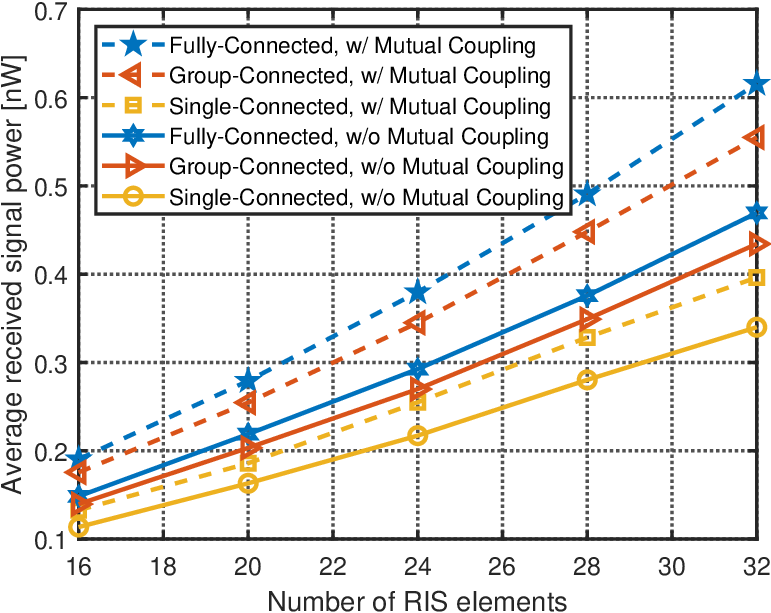}
\label{fig:case-studies-a}
}
\subfigure[$Y$- and $S$-parameters, w/o mutual coupling.]{
\includegraphics[width=0.4\textwidth]{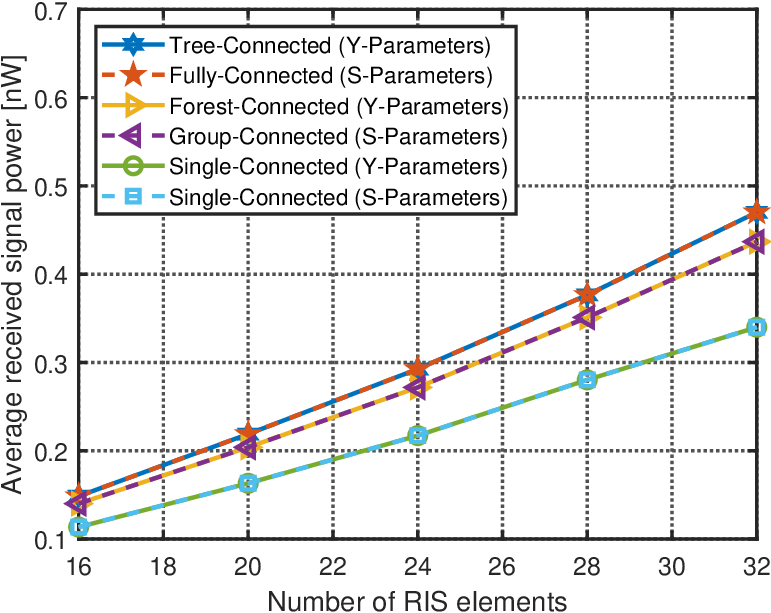}
\label{fig:case-studies-b}
}
\caption{Average received signal power maximized using $Z$-, $Y$-, and $S$-parameters.}
\label{fig:case-studies}
\end{figure*}

As a case study, we consider the received signal power maximization problem in an RIS-aided system, with the RIS being group-connected.
Considering lossless and reciprocal RIS, i.e., with $\boldsymbol{\Theta}^{H}\boldsymbol{\Theta}=\mathbf{I}$ and $\boldsymbol{\Theta}=\boldsymbol{\Theta}^T$, and $\mathbf{H}$ given in \eqref{eq:Hsimp-s}, assuming no mutual coupling, our problem writes as
\begin{align}
\underset{\mathbf{w},\mathbf{g},\boldsymbol{\Theta}}{\mathsf{\mathrm{max}}}\;\;
&P_{T}\left|\mathbf{g}\left(\mathbf{S}_{RT}+\mathbf{S}_{RI}\boldsymbol{\Theta}\mathbf{S}_{IT}\right)\mathbf{w}\right|^{2}\label{eq:PR-s}\\
\mathsf{\mathrm{s.t.}}\;\;\;
&\boldsymbol{\Theta}=\mathrm{diag}\left(\boldsymbol{\Theta}_{1},\ldots,\boldsymbol{\Theta}_{G}\right),\label{eq:PR-s-c1}\\
&\boldsymbol{\Theta}_{g}=\boldsymbol{\Theta}_{g}^{T},\;\boldsymbol{\Theta}_{g}^{H}\boldsymbol{\Theta}_{g}=\mathbf{I},\;\forall g,\label{eq:PR-s-c2}\\
&\left\|\mathbf{w}\right\|=1,\;\left\|\mathbf{g}\right\|=1,
\end{align}
which is solved by jointly optimizing $\mathbf{w}$, $\mathbf{g}$, and $\boldsymbol{\Theta}$.
To this end, we initialize $\boldsymbol{\Theta}$ to a feasible value and alternate between the following two steps.
First, with $\boldsymbol{\Theta}$ fixed, $\mathbf{w}$ and $\mathbf{g}$ are updated as the dominant right and left singular vectors of $\mathbf{S}_{RT}+\mathbf{S}_{RI}\boldsymbol{\Theta}\mathbf{S}_{IT}$, respectively, which is globally optimal.
Second, with $\mathbf{w}$ and $\mathbf{g}$ fixed, $\boldsymbol{\Theta}$ is updated by solving
\begin{equation}
\underset{\boldsymbol{\Theta}}{\mathsf{\mathrm{max}}}\;
\left\vert s_{RT}^{\text{eff}}+\mathbf{s}_{RI}^{\text{eff}}\boldsymbol{\Theta}\mathbf{s}_{IT}^{\text{eff}}\right\vert^{2}\;
\mathsf{\mathrm{s.t.}}\;
\eqref{eq:PR-s-c1},\;\eqref{eq:PR-s-c2},\label{eq:PR-s2}
\end{equation}
where $s_{RT}^{\text{eff}}=\mathbf{g}\mathbf{S}_{RT}\mathbf{w}$, $\mathbf{s}_{RI}^{\text{eff}}=\mathbf{g}\mathbf{S}_{RI}$, and $\mathbf{s}_{IT}^{\text{eff}}=\mathbf{S}_{IT}\mathbf{w}$.
Remarkably, \eqref{eq:PR-s2} has been solved for group-connected RISs through a closed-form global optimally solution given in \cite{ner22}.
These two steps are alternatively repeated until convergence of the objective function in \eqref{eq:PR-s}.

\section{Numerical Results}
\label{sec:results}

In this section, we evaluate the performance obtained by solving the three optimization problems presented in Sec.~\ref{sec:architecture-optimization}, using the $Z$-, $Y$-, and $S$-parameters.
The transmitter, RIS, and receiver are located at $(0,0)$, $(50,2)$, and $(52,0)$ meters (m), respectively.
We set $N_T=2$ and $N_R=2$, and assume that the direct channel between the transmitter and receiver is completely obstructed, i.e., $\mathbf{Z}_{RT}=\mathbf{0}$.
For the large-scale path loss of the channels from the RIS to the receiver and from the transmitter to the RIS, we use the distance-dependent path loss model $L_{ij}(d_{ij})=L_{0}d_{ij}^{-\alpha_{ij}}$, where $L_{0}$ is the reference path loss at distance 1~m, $d_{ij}$ is the distance, and $\alpha_{ij}$ is the path loss exponent, for $ij\in\{RI,IT\}$.
We set $L_{0}=-30$~dB, $\alpha_{RI}=2.8$, $\alpha_{IT}=2$, and $P_{T}=10$~mW.
For the small-scale fading, we model the channels as \gls{iid} Rayleigh, i.e., $\mathbf{S}_{RI}\sim\mathcal{CN}(\mathbf{0},L_{RI}\mathbf{I})$ and $\mathbf{S}_{IT}\sim\mathcal{CN}(\mathbf{0},L_{IT}\mathbf{I})$.
Given $\mathbf{Z}_{RT}$, $\mathbf{S}_{RI}$, and $\mathbf{S}_{IT}$, we obtain the rest of the off-diagonal blocks of $\mathbf{Z}$, $\mathbf{Y}$, and $\mathbf{S}$ through \eqref{eq:YRI-YIT}, \eqref{eq:YRT}, \eqref{eq:SRI-SIT}, and \eqref{eq:SRT}.

In Fig.~\ref{fig:case-studies}, we report the average received signal power obtained by optimizing the RIS through the $Z$-, $Y$-, and $S$-parameters.
In the case of group- and forest-connected RISs, the group size is $N_G=4$.
The schemes ``w/o Mutual Coupling'' are obtained by setting $\mathbf{Z}_{II}=Z_0\mathbf{I}$.
Besides, the schemes ``w/ Mutual Coupling'' in Fig.~\ref{fig:case-studies-a} are obtained by modeling $\mathbf{Z}_{II}$ as follows: its diagonal entries are set to $Z_0$, assuming perfectly matched RIS elements, while its off-diagonal entries are modeled as in \cite{li23-3}, considering the RIS antennas being dipoles with length $\ell=\lambda/4$ and inter-element distance $d=\lambda/4$, where $\lambda=c/f$ is the wavelength with frequency $f=28$~GHz.
We make the following observations.

\textit{First}, BD-RIS outperforms single-connected RIS in the presence of mutual coupling because of its higher flexibility, in agreement with \cite{li23-3}, and also in the absence of mutual coupling since Rayleigh fading is considered \cite{she20,ner23-1}.

\textit{Second}, the presence of mutual coupling between the RIS elements increases the performance, in accordance with \cite{akr23}.
We observe this gain for all the considered RIS architectures by assuming that the RIS mutual coupling matrix $\mathbf{Z}_{II}$ is perfectly known during the optimization process.

\textit{Third}, in the absence of mutual coupling, single-/group-/fully-connected RIS architectures optimized based on the $Z$-parameters achieve the same performance as the same architectures optimized based on the $S$-parameters, providing additional evidence of the equivalence of the two analyses.
In the considered case study, optimizing based on the $S$-parameters is preferred since closed-form solutions are available, leading to a low-complexity optimization process \cite{ner22}.

\textit{Fourth}, in the absence of mutual coupling, single-/group-/fully-connected RISs optimized by using the $S$-parameters achieve the same performance as single-/forest-/tree-connected RISs optimized with the $Y$-parameters, respectively, in agreement with \cite{ner23-1}.
In the considered case study, forest-/tree-connected RISs are preferred over group-/fully-connected RISs, since they are characterized by a reduced circuit complexity \cite{ner23-1}.

\section{Conclusion}
\label{sec:conclusion}

We introduced a universal framework to perform multiport network analysis of an RIS-aided communication system.
The proposed framework is used to analyze the RIS-aided system based on the $Z$-, $Y$-, and $S$-parameters.
Based on these three independent analyses, three equivalent channel models are derived accounting for the effects of impedance mismatching and mutual coupling at the transmitter, RIS, and receiver.
Subsequently, to gain insights into the role of the RIS in the communication model, these models are simplified by assuming large transmission distances, perfect matching, and no mutual coupling.
The equivalence between the obtained simplified models is shown by providing the mappings between the different parameters of the three representations.
The derived simplified channel model is consistent with the channel model widely used in related literature.
However, we show that an additional approximation is commonly considered in related literature, whose impact in terms of received signal power vanishes as the number of RIS elements increases but is non-negligible for a practical number of RIS elements.

Since the $Z$-, $Y$-, and $S$-parameters are equivalent representations, we discuss the advantages of each of them in the characterization of RIS architectures and their optimization.
To this end, we present three case studies and show that it is convenient to solve each of them by using a different multiport network model representation.
Numerical results further support the equivalence of the three analyses.

We identify three research directions opened by the physically consistent RIS-aided channel models presented in this study.
First, these models can be used to design RIS optimization algorithms aware of imperfect matching and mutual coupling, which are expected to lead to better performance gains given the more accurate channel modeling.
Second, our analysis allows for a more realistic assessment of the performance of RIS-aided systems, which can significantly improve the effectiveness of RIS deployment in realistic scenarios.
Third, the developed analysis based on the $S$-parameters can be used to simplify the experimental validation of RIS-aided channels through \glspl{vna}.
The RIS scattering matrix can be discriminated by leveraging the physics-consistent models developed in this study jointly with superposition and reciprocity principles.
Specifically, exploiting these principles, a \gls{vna} could be employed to separately assess the channels between the transmitter and RIS and between the RIS and receiver.
This circumvents the need for measuring the cascaded channel, a task often complicated due to the low \gls{snr}.

\bibliographystyle{IEEEtran}
\bibliography{IEEEabrv,main}

\end{document}